%% file: letter.tex
\documentclass[aps,prl,reprint, amsmath, amssymb,floatfix,superscriptaddress]{revtex4-2}
 
\usepackage{bm}
\usepackage{bbm}
\usepackage[retainorgcmds]{IEEEtrantools}
\usepackage{graphicx}
\usepackage{mathrsfs}
\usepackage{amsmath}
\usepackage{amsfonts}
\usepackage{amssymb}
\usepackage{color,xcolor}
\usepackage{times,txfonts}
\usepackage{nicefrac}
\usepackage{ragged2e}
\usepackage{float}
\usepackage{braket}

\usepackage{enumitem}
\usepackage{standalone}
\usepackage{tikz}
\usetikzlibrary{decorations.markings, arrows, fadings, backgrounds, calc}

\usepackage[italicdiff]{physics}

\usepackage[colorlinks=true,linkcolor=blue,urlcolor=blue,citecolor=blue,pdfusetitle]{hyperref}

\usepackage{blindtext}

\makeatletter
\DeclareRobustCommand{\cev}[1]{{\mathpalette\do@cev{#1}}}
\newcommand{\do@cev}[2]{%
  \vbox{\offinterlineskip
    \sbox\z@{$\m@th#1 x$}%
    \ialign{##\cr
      \hidewidth\reflectbox{$\m@th#1\vec{}\mkern4mu$}\hidewidth\cr
      \noalign{\kern-\ht\z@}
      $\m@th#1#2$\cr
    }%
  }%
}
\makeatother

\begin{document}

\title{Thermodynamic inference from excursions into hidden states: theory and quantum-dot experiment}
\date{\today}
\author{Guilherme Fiusa}
\affiliation{Department of Physics and Astronomy, University of Rochester, Rochester, New York 14627, USA}
\affiliation{University of Rochester Center for Coherence and Quantum Science, Rochester, New York 14627, USA}
\author{Jann van der Meer}
\affiliation{Department of Physics No. 1, Graduate School of Science, Kyoto University, Kyoto 606-8502, Japan}
\author{Tan Van Vu}
\affiliation{Center for Gravitational Physics and Quantum Information,
Yukawa Institute for
Theoretical Physics,\\ Kyoto University, Kitashirakawa Oiwakecho, Sakyo-ku, Kyoto 606-8502, Japan}
\author{Vivek Wadhia}
\affiliation{Department of Engineering Science, University of Oxford, Parks Road, Oxford OX1 3PJ, United Kingdom}
\author{Natalia Ares}
\affiliation{Department of Engineering Science, University of Oxford, Parks Road, Oxford OX1 3PJ, United Kingdom}
\author{Gabriel T. Landi}
\affiliation{Department of Physics and Astronomy, University of Rochester, Rochester, New York 14627, USA}
\affiliation{University of Rochester Center for Coherence and Quantum Science, Rochester, New York 14627, USA}
\author{Pedro E. Harunari}
\affiliation{Aix Marseille Université, CNRS, CINAM, Turing Center for Living Systems, 13288 Marseille, France}

\begin{abstract}

Inferring thermodynamic quantities from partially accessible data is a central problem in stochastic thermodynamics, where one is interested in what can be learned when only a part of the system is visible. 
We introduce an inference paradigm,
where an observer resolves a single
region of a Markov jump process and treats the rest as a black box, in which transitions are blurred or missed, and states are unresolved or hidden.
Each time the system leaves the visible region and returns, performing an excursion, the observer records only a time-antisymmetric counting observable and the excursion duration.
From the joint distribution of these two numbers we obtain a symmetry required by detailed balance, whose violation certifies that the hidden region is out of equilibrium, a lower bound on the largest affinity among cycles that are never directly observed, and an entropy production estimator. We apply all three results to a time trace showing the charge occupation of a double quantum dot in a cotunneling regime. We detect a violation of detailed balance within the first
$1\%$ of the record, and both bounds come close to the values of a generator fitted to the full trace, as expected when a single cycle dominates.

\end{abstract}

\maketitle{}

{\emph{Introduction}}---Stochastic thermodynamics has emerged as a comprehensive framework for characterizing nonequilibrium small systems. In contrast to macroscopic systems, the fluctuations of quantities such as heat, work, and entropy production are central~\cite{Jarzynski1997, Crooks1999, seifertEntropyProductionStochastic2005a, Seifert2012, Jarzynski2011, Shiraishi2023, Seifert2025book}. The most important feature of this framework is that thermodynamic laws can be formulated at the level of individual stochastic trajectories, providing fine-grained access to dissipation and irreversibility. The main issue, however, is that such characterization requires, in principle, fully resolving the microscopic dynamics. In practice, most real systems are only partially accessible, either by fundamental limitations, or experimental constraints such as finite sensitivity or time resolution~\cite{gingrich2026}. 

This question has motivated the rapidly growing field of thermodynamic inference~\cite{seifert2025universalboundsentropyproduction}, which has largely been organized around the different kinds of partial access an observer has to the underlying microscopic details. 
For discrete systems, several important classes of inference techniques for nonequilibrium and, more quantitatively, bounds on the rate of dissipation have been proposed that utilize, e.g., the thermodynamic uncertainty relation (TUR)~\cite{Barato2015, Gingrich2016, Horowitz2019, Hasegawa2019b}, 
waiting-time distributions~\cite{Berezhkovskii2019, Mart_nez2019, skinner2021, harunariWhatLearnFew2022, JannPRX2022, Jann2023, harunari2024periodic, pietzonka2024} and
correlations or correlation times~\cite{dechant2021, dechant2023, Ohga2023, raux2026lowerboundsentropyproduction}. These approaches assume a specific structure of the accessible data, and consequently, provide bounds that become tight under different circumstances. A major class of estimators for entropy production relies on extracting the irreversibility contained in a time series \cite{Gomez_Marin2008, roldan2010}. Depending on what an observer can resolve in a coarse-grained model, these methods can be specified to state lumping \cite{Puglisi2010, espositoStochasticThermodynamicsCoarse2012, bo2014, Teza2020}, transition-based models~\cite{harunariWhatLearnFew2022, JannPRX2022, harunari2023beat, garilliFluctuationRelationsFew2024a, ertel2024, harunari2024multifilar, Maier2025, Gu2026}, Markovian events \cite{Jann2023, JannaPRR2024} and higher-order (semi-)Markov processes \cite{wang2007, Jann2022, blom2024, schwarz2026}. 
The majority of the approaches focuses on estimating the rate of dissipation, but some works also establish bounds on the thermodynamic forces driving the system out of equilibrium~\cite{JannPRX2022, Ohga2023, Liang2023, JannaPRR2024}.

\begin{figure}\centering   
  \includegraphics[width=0.99\linewidth]{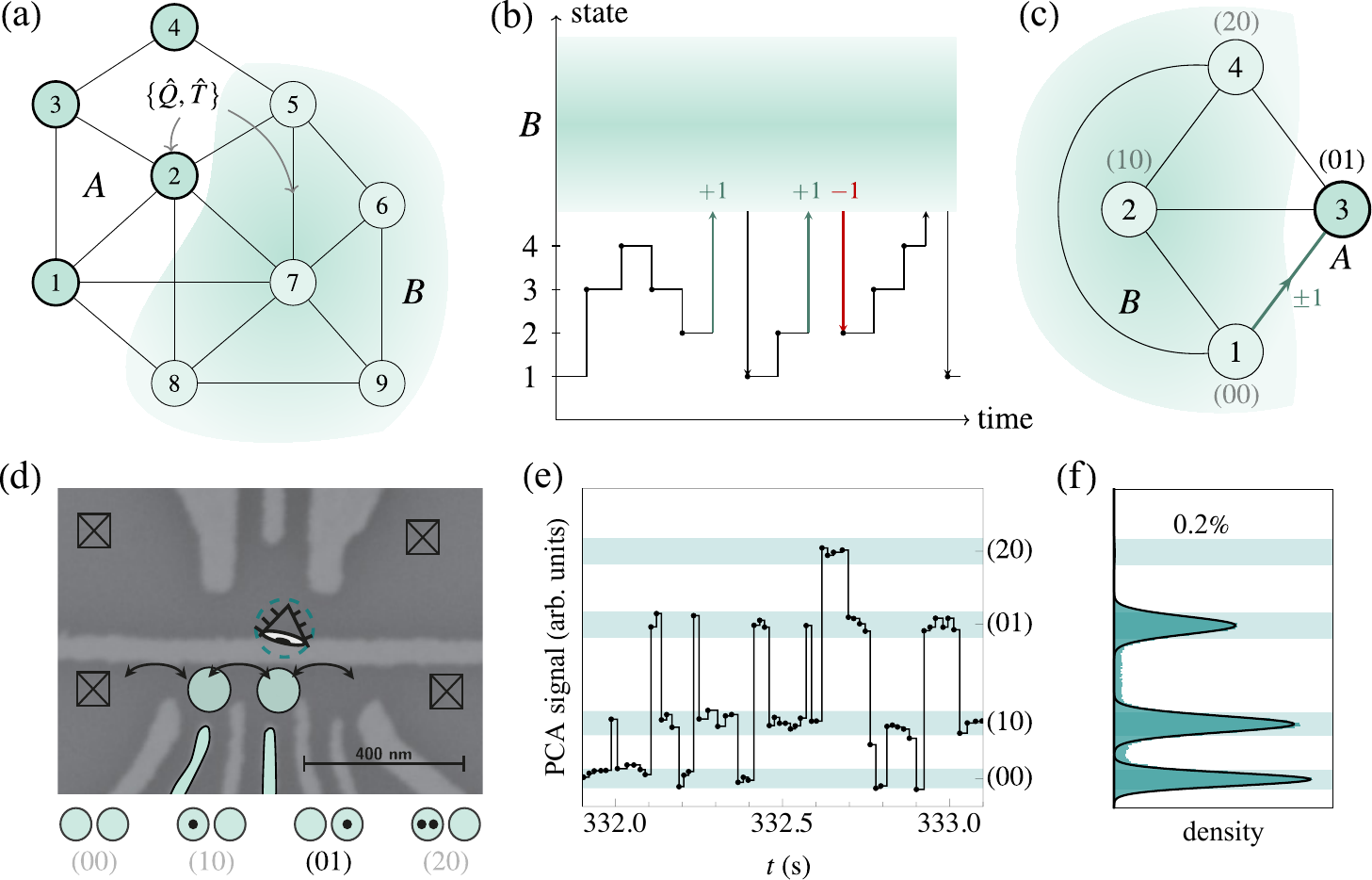}
\caption{(a) Generic Markov network with four visible states in region $A$ and a hidden region $B$ that may have arbitrarily many states.
Each excursion yields an outcome of the counting observable $\hat{Q}$ and a duration $\hat{T}$.
(b) Sketch of a trajectory following the topology in (a). 
As an example, the counting observable $\hat{Q}$ increases by $1$ whenever the system leaves state $2$ to enter $B$, and decreases by $1$ whenever it returns to state $2$ from $B$; other transitions do not contribute.
(c) Markov network of the four charge states of the double quantum dot explored in this work: $1=(00)$, $2=(10)$, $3=(01)$, and $4=(20)$ at positive bias [$(02)$ at negative bias]. The observer monitors only state $3$, in which the right dot holds exactly one electron, so that $A=\{3\}$ and $B=\{1,2,4\}$, and counts $+1$ for each jump $1\to3$ and $-1$ for each jump $3\to1$, i.e. an electron crossing the right barrier while the left dot is empty.
(d) Scanning electron micrograph of a device similar to the one used in this experiment. Green disks mark the left and right dots, formed under the highlighted plunger gates; arrows indicate tunneling between the reservoirs and the dots, and the eye marks the charge-sensor dot.
Below: the four charge configurations, with filled disks representing electrons.
(e) Short time window of the experimental charge-sensor trace~\cite{supp}; shaded regions mark the four charge levels.
(f) Histogram of the signal over the whole record with the Gaussian components of a fit that also accounts for the samples recorded during jumps~\cite{supp}. State $4$ does not form a separate peak, and accounts for only $0.2\%$ of the samples.}
\label{fig:markovian-network}
\end{figure}

However, in many settings the accessible data are organized differently: one can only tell whether the system resides in some visible region $A$, and record a single quantity each time the system returns to it, with no information on what happened in between. 
In other words, the system visits a black box region $B$ for a random amount of time. This is what we call an excursion over $B$.
The recorded quantity is a counting observable $\hat{Q}$, which projects the possibly complicated dynamics in $B$ onto a single number per excursion [see Fig.~\ref{fig:markovian-network}].
Approaches on the level of the stochastic trajectory, such as the coarse-grained entropy production~\cite{JannaPRR2024} or the excursion framework~\cite{Fiusa2025, Fiusa2025a}, allow us to study the fluctuations of entropy production and other observables over a single excursion, rather than over a fixed time interval.

In this Letter, we introduce a complementary inference paradigm for any continuous-time Markov jump process, based on time-antisymmetric counting observables (such as currents, entropy production, energy exchanges) over stochastic excursions. The observer's data is the joint distribution of outcomes and durations $P(q,t)$,
and we ask what can be inferred about the black box region $B$ from counting observables and excursion durations alone.
We derive three results. 
First, detailed balance forces $P(q,t) = P(-q,t)$, which yields two independent tests: an equality of outcome frequencies and an equality of conditional duration distributions.
Second, the log-ratio $\ln \left[P(q)/P(-q)\right]$ lower-bounds the largest affinity among cycles carrying outcome $q$ in the underlying network, giving access from partial data to the nonequilibrium forces that drive the hidden region.
Third, we derive a lower bound on the total entropy production rate $\hat{\sigma}_{\rm exc} \le \sigma$. The latter two results saturate for unicyclic
networks~\cite{supp}.

The excursion is the natural unit of observation in several systems, from cellular sensing~\cite{Harvey2023} to experimentally realized quantum clocks~\cite{Wadhia2025}, heat engines~\cite{Josefsson2018}, and absorption refrigerators~\cite{aamirThermallyDrivenQuantum2025}.
In double quantum dots, excursions were used to characterize charge transport, its fluctuations, and the precision trade-offs that govern it~\cite{fiusa2026stochastictrajectoriesexcursionsdouble}.
With that in mind, we go beyond a formal demonstration and consider strategies to overcome challenges of real data. We employ statistical tests and bootstrapping, obtain $p$-values, and consider effects of finite sampling rate. Equipped with our results, we uncover properties of a double quantum dot (DQD) device operated in a cotunneling regime that are hidden from an observer monitoring a single dot. Counting excursion outcomes, with no model of the hidden states, reveals the violation of detailed balance at the $10^{-3}$ level within the first $1\%$ of the record. With $10\%$ of the record, the lower bounds on the cycle affinity and entropy production already lie within the $95\%$ intervals of their full-record values. 
The same analysis singles out a dissipative channel through a charge configuration too rarely occupied to be resolved in the signal itself.

{\emph{Setting and counting observables}}---We consider a Markov jump process, described by the master equation 
\begin{equation}\label{M}
\frac{d|p\rangle}{dt} = W|p\rangle,
\end{equation} where $|p\rangle$ is the probability vector with entries $p_x$, i.e., the
probability of finding the system in a state $x$, and $W$ is the generator, whose off-diagonal entries $W_{xy}$ are the rates of the transitions $y\to x$ and whose diagonal entries are $W_{xx}=-\Gamma_x$, with $\Gamma_x=\sum_{y\,(\neq x)}W_{yx}$ the escape rate from $x$.
When $W$ is irreducible, there is a unique normalized $|\pi \rangle $ that is the solution of $d|\pi \rangle /dt = 0$. Throughout this work, we assume that the system is in this steady state.
We now split the state space into two regions, $A$ and $B$, and consider trajectory segments that begin with a transition $A \to B$ and end with the next transition back $B \to A$, i.e., excursions over region $B$.

The main ingredient introduced by Refs.~\cite{Fiusa2025, Fiusa2025a} in stochastic excursions is the idea of counting observables, which is defined as
\begin{equation}\label{counting_observable}
    \hat{Q} = \sum_{x,y} \nu_{xy} \hat{N}_{xy},
\end{equation}where $\hat{N}_{xy}$ is a random variable that records the number of transitions from $y$ to $x$ in an excursion, and $\nu_{xy}$ are generic weights that give physical meaning to the transitions.
The counting observable $\hat{Q}$ may include the first transition $A\to B$, transitions within $B$, and the last transition $B \to A$. 

In this Letter, we consider time-antisymmetric observables, i.e. those whose weights obey $\nu_{xy} = -\nu_{yx}$~\footnote{Thermodynamic observables such as heat and entropy production take this form. Those observables are usually called currents.}, 
and enhance the excursion formalism by providing tools to do inference over region $B$. 
Whatever the observer can fully resolve becomes region $A$, while region $B$ is a black box: transitions can be blurred or missed, states are lumped or unresolved [Fig.~\ref{fig:markovian-network}(a),(b)]. 
Here, we assume that, despite these limitations, the observer records per excursion the realization of a time-antisymmetric counting observable and the excursion duration.
The observer's data is thus the joint distribution $P(q,t)$ of the outcome $q$ and the duration $t$, accumulated over many excursions. 
We show what can be inferred about region $B$ from the joint distribution.
All inference results developed in this Letter are applied to experimental trajectory data from a DQD setup in a cotunneling regime [Fig.~\ref{fig:markovian-network}(d)--(f)], and compared to results unveiled by a generator reconstructed from the full trace, which is not available to an observer of $A$ alone.

{\emph{Detailed balance and excursion durations}}---The first question to address in our framework is whether detailed balance (DB) in the region $B$ can be detected from excursion data alone. DB means that every stationary probability current vanishes, $W_{xy}\pi_y = W_{yx}\pi_x$ for all pairs of states $x, y$. Verifying it directly requires resolving all transition rates, which is not possible under the inference paradigm. 
We show here that excursions nonetheless carry a necessary condition for DB, for any time-antisymmetric counting observable.

Under DB, the path weight of an excursion equals that of its time reverse up to the ratio of the steady state occupations of the two visible states at which it begins and ends; the contributions of all the in-between states in $B$ cancel out~\cite{supp}. This ratio is one whenever the excursion begins and ends at the same state, and in particular for every excursion when $|A| = 1$.
Since time reversal flips the sign of $\hat{Q}$ and preserves the duration, DB implies $P(q,t)=P(-q,t)$, which provides a symmetry of frequency and duration statistics for a given outcome of the observable, i.e.
\begin{equation}
\label{eq:DB-symmetry}
  P(q) = P(-q)
  \qquad\text{and}\qquad
    \dfrac{P(t \ |\ q)}{P(t \ | -q)} = 1,
\end{equation}where $q$ and $-q$ are two (of potentially many) outcomes of the counting observable $\hat{Q}$. 
For $|A|>1$, the same argument holds at the level of excursions resolved by the visible states through which the system leaves and returns to $A$ (see End Matter).

Both equalities~\eqref{eq:DB-symmetry} are falsifiable with empirical statistics over excursions, and the failure of either implies that DB is violated in region $B$.
 The outcome frequencies give the simpler test, requiring only 
the number of excursions that carry each outcome $q$; the duration statistics give a complementary one, sensitive to asymmetries that survive even when the frequencies happen to balance. 
A violation implies sustained probability currents in $B$, none of which is directly observed.

\begin{figure*}
    \centering
    \includegraphics[width=0.99\linewidth]{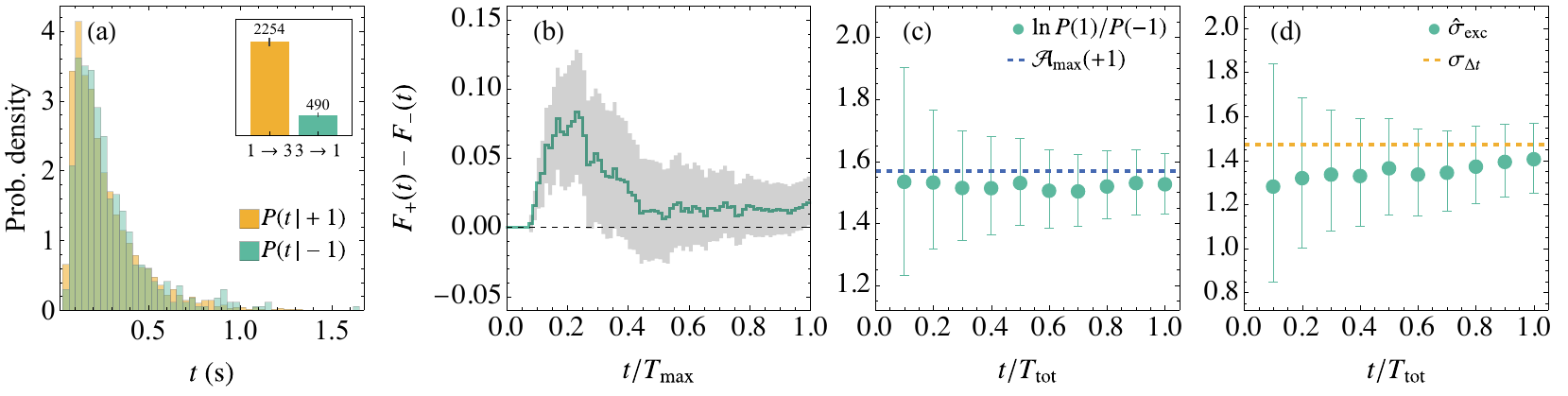}
    \caption{Results of this Letter applied to the double quantum dot at positive bias $0.855$\,V, with region $A=\{3\}$ (single occupation of the right dot) and the counting observable that records $\pm1$ for the jumps $1\leftrightarrow3$ [Fig.~\ref{fig:markovian-network}(c)].
    (a)-(b): Test of the detailed balance condition in Eq.~\eqref{eq:DB-symmetry}.
    (a) Empirical distributions $P(t\,|+1)$ and $P(t\,|-1)$ of the excursion durations. \emph{Inset}: number of excursions with $q=+1$ ($n_+=2254$) and $q=-1$ ($n_-=490$), with $95\%$ Poisson intervals.
    (b) Difference $F_+(t)-F_-(t)$ of the corresponding cumulative distributions, against $t/T_{\max}$ with $T_{\max}$ the $98\%$ quantile of the durations. The line is the data and the shaded region is a $95\%$ bootstrap band. The Kolmogorov--Smirnov distance $D=\max_t|F_+(t)-F_-(t)|\simeq0.09$ has permutation $p$-value $4.5\times10^{-3}$.
    (c) $\ln[P(+1)/P(-1)]$ computed from cumulative time fractions $t/T_{\rm tot}$ of the total time record, with $95\%$ bootstrap intervals, against $\mathcal{A}_{\max}(+1)\simeq1.57$ (dashed), obtained independently from the generator fitted to the same trace~\cite{supp}.
    (d) Excursion estimator $\hat{\sigma}_{\rm exc}$, Eq.~\eqref{eq:bound-ep}, from the same cumulative time fractions of the record, against $\sigma_{\Delta t}\simeq1.47\,\mathrm{s}^{-1}$ (dashed), the entropy production rate of the generator fitted to the same sampled trace~\cite{supp}. Both estimators lie close to and below their benchmarks.}
    \label{fig:db-inference}
\end{figure*}

{\emph{Affinity bound}}---A sequence of transitions that forms a closed loop without self-crossings is defined as a cycle $c$. 
Forces may drive the cycle in a particular orientation. This driving is quantified through the thermodynamic affinities, defined as \cite{Hill1966, schnakenberg1976, Hill1989}
\begin{equation}\label{eq:affinity}
    \mathcal{A}_{c} := \ln{\prod_{(x,y) \in c}\dfrac{W_{xy}}{W_{yx}}} = \ln{\prod_{(x,y) \in c}\dfrac{W_{xy} \pi_y}{W_{yx} \pi_x}}.
\end{equation}
Under local detailed balance \cite{seifert2018, maes2021}, $\mathcal{A}_c$ is identified as the entropy produced in the environment per completion of the cycle $c$. 
A nonvanishing $\mathcal{A}_{c} \neq 0$ for a single cycle $c$ suffices to imply broken DB, but unless such a cycle $c$ is fully contained in the visible region $A$, it is not possible to infer nonequilibrium directly from Eq.~\eqref{eq:affinity}. 

A counting observable $\hat{Q}$ in region $B$ can encode information about such hidden cycles. We show here that for $|A| = 1$ the ratio of distributions of its outcomes (cf. Eq. \eqref{eq:DB-symmetry}) provides a lower bound on the maximum affinity $\mathcal{A}_\text{max}$ in the network of the form
\begin{equation}
\label{eq:bound-max-affinity}
    \ln \frac{P(q)}{P(-q)} \leq \mathcal{A}_\text{max} (q) \leq \mathcal{A}_\text{max},
\end{equation}
where $\mathcal{A}_\text{max}(q) = \max_{c : \hat{Q}(c) = q} \mathcal{A}_c$. The bound requires meeting one of the following sufficient conditions:

\begin{enumerate}[label=Condition \arabic*:, leftmargin=*, topsep=2pt, itemsep=0pt, parsep=2pt]
    \item There are no cycles fully contained in $B$, i.e., any cycle $c$ passes through the state of $A$.
    \item $\hat{Q}$ has increments only at the boundary between $A$ and $B$, i.e., $\nu_{xy}=0$ when both $x,\ y \in B$.
\end{enumerate}
Condition 1 permits arbitrary time-antisymmetric counting observables, whereas Condition 2 does not restrict topology. The End Matter explains why Eq. \eqref{eq:bound-max-affinity} can be violated without any such additional conditions and contains the derivation from Condition 1. We refer to the Supplemental Material \cite{supp} for the proof from Condition 2. 

For $|A| > 1$, an excursion does not give rise to a closed cycle, unless it leaves and enters $A$ through the same state, in which case the bound \eqref{eq:bound-max-affinity} remains valid. The general case is discussed briefly in the End Matter and in detail in the Supplemental Material \cite{supp}.

{\emph{Entropy production bound}}---The steady-state entropy production rate is defined as \cite{schnakenberg1976}
\begin{equation}
    \sigma = \sum_{x,y} W_{xy} \pi_y \ln \frac{W_{xy} \pi_y}{W_{yx} \pi_x},
\end{equation}
in units of Boltzmann constant. Assessing its value is usually unfeasible since it requires resolving every state and transition.
Here we provide an estimator built directly from the distribution of excursion observables $P(q)$. 
The result reads
\begin{equation}
\label{eq:bound-ep}
\hat{\sigma}_{\rm exc} :=
   K\sum_q P(q) \ln{\left(\dfrac{P(q)}{P(-q)}\right)}  \leq \sigma,
\end{equation}
where $K$ denotes the number of excursions per unit time (excursion rate). 
This is valid for $|A|=1$, but we discuss the general case in the End Matter, and provide the proofs in \cite{supp}. Moreover, we also show in \cite{supp} that the bound saturates in unicyclic networks, and vanishes precisely when the outcome distribution is symmetric, in accordance with Eq.~\eqref{eq:DB-symmetry}.

Two main features distinguish Eq.~\eqref{eq:bound-ep} from existing estimators. First, it recovers part of the visible-transition bound of Refs.~\cite{JannPRX2022, harunariWhatLearnFew2022} as a particular case: choosing weights $\nu_{xy} = -\nu_{yx} = 1$ for a particular $x \leftrightarrow y$ transition and all other transitions to zero. Observing a transition $y \to x$ amounts to one excursion with outcome $\hat{Q} = +1$, provided the excursion returns through a channel of vanishing weight.
Second, Eq.~\eqref{eq:bound-ep} does not require resolving individual forward and backward transitions. 
Since $A$ is always visible, we can identify whether a transition leaves or returns to $A$ without specifying which edge was traversed.
This has a direct consequence for how much statistics a given trace
provides.
The estimators of Refs.~\cite{JannPRX2022,
harunariWhatLearnFew2022} are built from the waiting time between two consecutive forward transitions $y \to x$, and between two consecutive backward ones $x \to y$, so each sample requires a pair of visible transitions, whereas the excursion estimator requires a single outcome of the counting observable. Depending on the topology, excursions therefore gather sampling points faster than transition-based approaches.

\emph{Double quantum dot experiment and challenges of real data}---We benchmark our results experimentally on a double quantum dot (DQD) device [Fig.~\ref{fig:markovian-network}(d)], with electrostatically defined dots on a Ge/SiGe heterostructure \cite{jirovec2021singlet} with gate electrodes patterned on top. This device was loaded into a dilution refrigerator and cooled to 180 mK.
A charge sensor dot is capacitively coupled to the DQD, and is used to monitor its charge occupation state. By applying a voltage across, and a radiofrequency (rf) signal to, the sensor dot, a signal is able to be measured that takes on distinct values corresponding to this occupation. Additionally, source and drain reservoirs allow us to apply a voltage bias across the DQD, which results in charge transport being favored in one direction. For more details about the experimental setup and measurement see \cite{supp}.

The charge-sensor trace contains four charge states: $1=(00)$ (both dots empty), $2=(10)$ and $3=(01)$ (a charge in the left or in the right dot), and $4=(20)$ at positive bias or $(02)$ at negative bias; the configuration $(11)$ is never occupied and enters only as the intermediate state of cotunneling processes [Fig.~\ref{fig:markovian-network}(c)--(f)].
In the sequential regime, charges tunnel one at a time between a dot and a lead, or between the two dots, and each jump changes the charge of one dot by one charge carrier. Cotunneling is the second-order process in which two such tunneling events happen in a single step~\cite{Geerligs1990, Franchesi_2001, van_der_Wiel2002, Rudge2018, Golovach2004, Pedersen2007, Koch2006}.

We emulate an observer who monitors only the right dot and resolves whether it holds exactly one charge. The visible region is then $A=\{3\}$, whereas $B=\{1,2,4\}$.
As counting observable we take the net number of charges that cross the right barrier, counting $+1$ for a jump $1 \to 3$ and $-1$ for $3 \to 1$.
At positive bias these are the only transitions through the right barrier, so $\hat{Q}$ is the charge exchanged between the right dot and its lead during an excursion.
This is a model assumption: the observer needs no access to the left dot, but must resolve the right barrier. Here we evaluate $\hat{Q}$ from the reconstructed trajectory.
Since each excursion contains exactly one transition out of $A$ and one back into $A$, the possible outcomes are $0$ and $\pm1$: $q=+1$ when the excursion leaves $3$ towards $2$ or $4$ and returns from $1$, $q=-1$ when it leaves towards $1$ and returns from $2$ or $4$, and $q=0$ otherwise.

We now apply the three results to the measured trace at a positive DQD bias of $0.855$\,V~\cite{data}. Over the full $1915$\,s record we identify $6424$ excursions, of which $2744$ carry a nonzero outcome, split as $n_{+}=2254$ for $q=+1$ against $n_{-}=490$ for $q=-1$. Under detailed balance the two outcomes are equally likely, and this split has two-sided binomial probability below $10^{-200}$, so the frequency equality in Eq.~\eqref{eq:DB-symmetry} is violated.
%Note that this test requires no more than counting how many excursions carry each outcome.
The duration statistics [also in Eq.~\eqref{eq:DB-symmetry}] provide an additional test. The Kolmogorov--Smirnov distance between the empirical conditionals $P(t\,|+1)$ and $P(t\,|-1)$ is $D\simeq0.09$ [Fig.~\ref{fig:db-inference}(a),(b)], with a permutation $p$-value of $4.5\times10^{-3}$ obtained by shuffling the outcome labels across excursions. The violation is detected with very little data: the binomial $p$-value computed on the beginning of the record falls below $10^{-3}$ after the first $12.5$\,s, less than $1 \%$ of the record ($19$ excursions with nonzero outcome), and below $10^{-5}$ after $21.1$\,s. Since these are first crossings of a test repeated as data accumulate, they indicate how much data is needed rather than serving as significance levels.

The same two counts $n_{\pm}$ give $\ln[P(+1)/P(-1)]\simeq1.53$, with a $95\%$ bootstrap interval $[1.43,1.63]$ and a conservative one-sided lower limit of $1.44$ at $95\%$ confidence~\cite{supp}. This is the quantity bounded by Eq.~\eqref{eq:bound-max-affinity}. The largest affinity among cycles with outcome $+1$, obtained independently from the generator fitted to the same trace, is $\mathcal{A}_{\max}(+1)\simeq1.57$, carried by the cycle $3\to2\to1\to3$~\cite{supp}.
The excursion rate is $K\simeq3.354\,\mathrm{s}^{-1}$.
Inserting $K$ into Eq.~\eqref{eq:bound-ep} gives $\hat{\sigma}_{\rm exc}\simeq1.41\,\mathrm{s}^{-1}$, with bootstrap interval $[1.25,1.57]\,\mathrm{s}^{-1}$ and conservative lower limit $1.21\,\mathrm{s}^{-1}$, against the entropy production rate $\sigma_{\Delta t}\simeq1.47\,\mathrm{s}^{-1}$ obtained from the generator fitted to the same trace~\cite{supp}.
From the first $10\%$ of the record onward [Fig.~\ref{fig:db-inference}(c),(d)], both point estimates already lie within the $95 \%$ intervals of their full-record values.
Both bounds lie close to their benchmarks: $\ln[P(+1)/P(-1)]$ reaches $97\%$ of $\mathcal{A}_{\max}(+1)$, and $\hat{\sigma}_{\rm exc}$ reaches $96\%$ of $\sigma_{\Delta t}$. 
Since state $4$ is rarely occupied, the dynamics is close to that of the ring $1$--$2$--$3$, a unicyclic network in which Eqs.~\eqref{eq:bound-max-affinity} and~\eqref{eq:bound-ep} saturate~\cite{supp}; the edges of this ring carry $99\%$ of $\sigma_{\Delta t}$. 
We also consider a time trace at negative bias ($-0.955$\,V) in ~\cite{supp}, where state $4$ is occupied only $0.07\%$ of the time and does not appear as a separate peak in the signal histogram, yet the excursion statistics single it out as the channel that carries the dissipation.

Practical applications involve challenges such as finite sampling rate and noise. These can hinder the accurate identification of excursions and the evaluation of their counting observables.
The finite sampling rate can be seen as a map $M$ taking the continuous trajectory $\gamma$ into the sparse $M \gamma$, whose points are equally spaced by $\Delta t$. Here $1/\Delta t \simeq 58.1\,\mathrm{s}^{-1}$ against escape rates of $10$--$26\,\mathrm{s}^{-1}$~\cite{supp}, so a residence is resolved by only $2$ to $6$ samples and transitions are genuinely missed: consecutive excursions may be read as one, and increments of the counting observable may be lost. 
Since $M$ commutes with time reversal, $\theta M\gamma = M\theta\gamma$, the log-sum inequality still gives $\hat{\sigma}_{\rm exc} \leq \sigma$ for the sampled trajectory. The benchmark is degraded by the same sampling: $\sigma_{\Delta t}$ is computed from a generator fitted to $M\gamma$, and on emulated records with known rates it lies below $\sigma$ in every case~\cite{supp}. What Fig.~\ref{fig:db-inference}(d) shows is therefore a comparison between two quantities built from the same sampled trace. A natural concern is whether such time coarse graining introduces memory.
At the level of the counting observable it does not: every sampled visit to $A$ is a renewal point, so sampled excursions, including those merged by a missed return, remain i.i.d., and the entropy production bound holds for them.
The same holds for the processing that turns the signal into a trajectory: each sample is assigned to the nearest level of a fit to the signal histogram~\cite{supp}, visits shorter than two samples are discarded, except for the short-lived state $4$, and repeated states are merged. All three steps commute with time reversal, so the bounds hold for the processed trajectory, and $\hat{\sigma}_{\rm exc}$ and $\sigma_{\Delta t}$ are compared on the same footing. Absolute values depend on these choices, whereas the near-saturation and the asymmetries reported above do not~\cite{supp}.

{\emph{Conclusions}}---We have introduced an inference paradigm for partially accessible Markov jump processes based on stochastic excursions, in which accessible data consists of the joint distribution $P(q,t)$ of a time-antisymmetric counting observable and the excursion duration. 
We showed that this visible data is enough to learn about the hidden region in three different ways: a necessary condition for detailed balance, a lower bound on maximal cycle affinities, and a lower bound on the total entropy production rate. 
%The estimators we derived for entropy production and affinities saturate on unicyclic networks.
We benchmarked the framework on a double quantum dot operated in a cotunneling regime, with an observer that monitors a single dot.
The violation of detailed balance is detected from a single visible state, a counting observable, and about $1\%$ of the total trace, without reconstructing the hidden dynamics.
Since long acquisitions in gate-defined devices are limited by slow drifts~\cite{Paladino2014, Massai2024, Stehouwer2025}, this makes the tests applicable within windows short enough for stationarity to hold, and opens the way to tracking how nonequilibrium properties change across a measurement.
At positive bias, about $10\%$ of the trace is sufficient to obtain point estimates of both bounds that lie within the $95\%$ intervals of their full-record values.
At negative bias, the excursion statistics single out a rare, directed channel through a state too seldom occupied to appear in the signal histogram.

{\emph{Acknowledgments}}---G.F. thanks Udo Seifert and his group for the great hospitality during an extended visit to the University of Stuttgart, and for numerous valuable discussions there.
P.E.H. and G.F. also thank Pierre Ronceray and João Valeriano for fruitful discussions. 
This research is primarily supported by the U.S. Department of Energy (DOE), Office of Science, Basic Energy Sciences (BES) under Award No. DE-SC0025516.
G.F. and J.v.d.M. thank the Yukawa Institute for Theoretical Physics at Kyoto University and MEXT KAKENHI Grant‑in‑Aid for Transformative Research Area (B) “Quantum Energy Innovation”, where this work was initiated during “Kyoto Workshop on Quantum Thermodynamics and Stochastic Thermodynamics 2025” (YITP-I-25-03). 
J.v.d.M. was supported by JSPS KAKENHI Grant No. 25KF0238 and JSPS Research Fellowship No. P25761.
T.V.V. was supported by JSPS KAKENHI Grants No.~JP23K13032, No.~JP26K00019, No.~JP26K00022, and No.~JP26H02015.
V.W. acknowledges funding from UK Research and Innovation grant EP/T517811/1. N.A. was supported by the European Research Council (grant agreement 948932) and the Royal Society (URF-R1-191150). The experiments were co-funded by the European Union (Quantum Flagship project ASPECTS, Grant Agreement No. 101080167) and UK Research and Innovation (UKRI). The views and opinions expressed are those of the authors only and do not necessarily reflect those of the European Union, the Research Executive Agency, or UKRI.
PH acknowledges funding from the European Union (ERC-SuperStoc-101117322).

{\emph{Data availability}}---The measurement traces as well as
the code for analysis is publicly available on GitHub~\cite{data}.

\appendix 
\bibliography{Letter}

\pagebreak

\onecolumngrid

\begin{center}
    \textbf{End Matter}
\end{center}
\twocolumngrid

\textit{General case} $|A| > 1$---All three results were stated for a single state in region $A$, which is the situation of the observer in the DQD experiment. We now discuss what happens when $A$ has several states. Throughout, $A$ is fully resolved, so the observer records for each excursion the state $x \in A$ from which it left and the state $y \in A$ to which it returned, together with the stationary occupations $\pi_x$ of the visible states and the rates $\pi_x K_{xy}$ of excursions from $x$ to $y$. All of these are accessible from the trace.

Let $P_{xy}(q,t)$ denote the joint distribution of outcome and duration of excursions that leave $A$ at $x$ and return at $y$. Under detailed balance, the ratio of path weights between an excursion and its reverse reduces to $\pi_y / \pi_x$, which is constant at fixed endpoints and therefore factors out of the conditioning~\cite{supp}. Two relations follow,
\begin{equation}
  \frac{P_{xy}(t \,| q)}{P_{yx}(t \,|-q)} = 1,
  \qquad
  \frac{P_{xy}(q)}{P_{yx}(-q)} = \frac{\pi_y}{\pi_x} \frac{K_{yx}}{K_{xy}},
  \label{eq:em-db-general}
\end{equation}which reduce to Eq.~\eqref{eq:DB-symmetry} when $|A| = 1$, since then $\pi_y/\pi_x = 1$ and $K_{xy} = K_{yx}$. The operational aspect of both results remains the same as the duration statement is unchanged in form, and the frequency statement acquires a prefactor built entirely from visible quantities. The failure of either certifies a violation of detailed balance in $B$.

\textit{Proof of the entropy production estimator}---Let $\gamma_t=(x_1,t_1,x_2,t_2,\dots)$ be a typical path that is generated by the Markov jump process for a long time $t$, where $x_1\in A$.
Here, the system stays at state $x_i$ for a duration $t_i$, then jumps to the next state $x_{i+1}$.
In general, $\gamma_t$ can be decomposed into a sequence of independent complete excursions, followed by an incomplete residual path at the end; that is, $\gamma_t=(\mathcal{E}_1,\dots,\mathcal{E}_{N_t},\mathcal{I})$.
Each excursion $\mathcal{E}_k$ is of the form $(x_1,t_1,\dots,x_{N-1},t_{N-1},x_N)$, where $x_1,x_N\in A$ and $x_i\in B$ for all $1<i<N$.
Let $\Delta s_k=\sum_{i=1}^{N-1}\ln(W_{x_{i+1}x_i}/W_{x_ix_{i+1}})$ denote heat dissipation along the excursion $\mathcal{E}_k$ and $\Delta s_{\rm res}$ be heat dissipation along the residual path $\mathcal{I}$.

Due to the ergodicity of Markov jump processes, the entropy production rate can be expressed as
\begin{equation}
    \sigma=\lim_{t\to\infty}\dfrac{1}{t}\ln\frac{P(\gamma_t)}{P(\tilde{\gamma}_t)},
\end{equation}
where $P(\gamma_t)=e^{-\Gamma_{x_1}t_1}W_{x_2x_1}e^{-\Gamma_{x_2}t_2}W_{x_3x_2}\dots$ and $\Gamma_{x} = \sum_{y(\neq x)}W_{yx}$ is the escape rate from $x$.
Since $N_t\to\infty$ as $t\to\infty$, the entropy production rate can be calculated as follows:
\begin{align}\label{eq:end-matter-sigma-decomp}
    \sigma&=\lim_{t\to\infty}\frac{\Delta s_{\rm res} + \sum_{k=1}^{N_t}\Delta s_k}{t}\notag\\
    &=\lim_{t\to\infty}\qty(\frac{\Delta s_{\rm res} + \sum_{k=1}^{N_t}\Delta s_k}{N_t})\qty(\frac{N_t}{t})\notag\\
    &=K\ev{\Delta s_k},
\end{align}
where $K=\lim_{t\to\infty}N_t/t$ denotes the rate of excursions.
For each excursion $\mathcal{E} = (x_1, t_1, \ldots, x_{N-1}, t_{N-1}, x_N)$, the path weight $P(\mathcal{E})$ is defined just like $P(\gamma_t)$ but with fixed endpoints of the excursion, collecting the $N-1$ jump factors and the residence factors of the interior states only, divided by the escape rate $\Gamma_{x_1}$, so that it is the probability of the excursion once the system leaves $x_1$.
Since the probability that the system starts from $A$ and returns to $A$ at some time is equal to $1$ i.e. $\sum_{\mathcal{E}}\mathcal{D}[\mathcal{E}]P(\mathcal{E})=1$, where
$\sum_{\mathcal{E}}\mathcal{D}[\mathcal{E}] \equiv \sum_{N\geq3}\sum_{x_2,\ldots,x_{N-1}\in B}\int_0^\infty dt_2\cdots\int_0^\infty dt_{N-1}$
runs over the number of visited states, the sequence of states visited in $B$,
and their residence times (the boundary residences in $A$ are not part of the
excursion), and is invariant under time reversal,
$\mathcal{D}[\mathcal{E}]=\mathcal{D}[\tilde{\mathcal{E}}]$, the average heat
dissipation can be evaluated as%$\sum_{\mathcal{E}}D[\mathcal{E}]P(\mathcal{E})=1$, where $\sum_{\mathcal{E}}D[\mathcal{E}]$ is the shorthand for the sum of all excursions, the average heat dissipation can be evaluated as
\begin{align}
    \ev{\Delta s_k}&=\sum_{\mathcal{E}}D[\mathcal{E}]P(\mathcal{E})\ln\frac{P(\mathcal{E})}{P(\tilde{\mathcal{E}})}\ge\sum_{q}P(q)\ln\frac{P(q)}{P(-q)},
\end{align}
where we applied the log-sum inequality to obtain the last inequality~\cite{cover2012elements}.
Consequently, we obtain the following lower bound for entropy production rate in the case $|A|=1$:
\begin{equation}\label{eq:app:bound-A-1}
    \sigma\ge KD_{\rm KL}(P(q)\|P(-q)).
\end{equation}where $D_{\rm KL}$ denotes the Kullback-Leibler divergence and all quantities in Eq.~\eqref{eq:app:bound-A-1} can be estimated from trajectory data alone.

A similar result also holds for $|A| > 1$, we show the proof in \cite{supp}. The main technicality is that now a typical path can no longer be decomposed into excursions only, i.e. it becomes interspersed with paths consisting of states belonging to $A$, followed by an incomplete residual path at the end; that is, $\gamma_t=(\mathcal{E}_1,F_1,\mathcal{E}_2,F_2,\dots,\mathcal{E}_{N_t},F_{N_t},\mathcal{I})$. In turn, following the same argument developed here, the decomposition in Eq.~\eqref{eq:end-matter-sigma-decomp} simply acquires one extra term due to the dissipation along the paths $F_k$:
\begin{equation}
   K\sum_q P(q) \ln{\left(\dfrac{P(q)}{P(-q)}\right)} +\sigma_A \leq \sigma,
\end{equation}
which takes into account the entropy production rate within region $A$, $\sigma_A$. Since $A$ is fully accessible, it can be directly evaluated with no need for inference methods.

\begin{figure}[ht!]
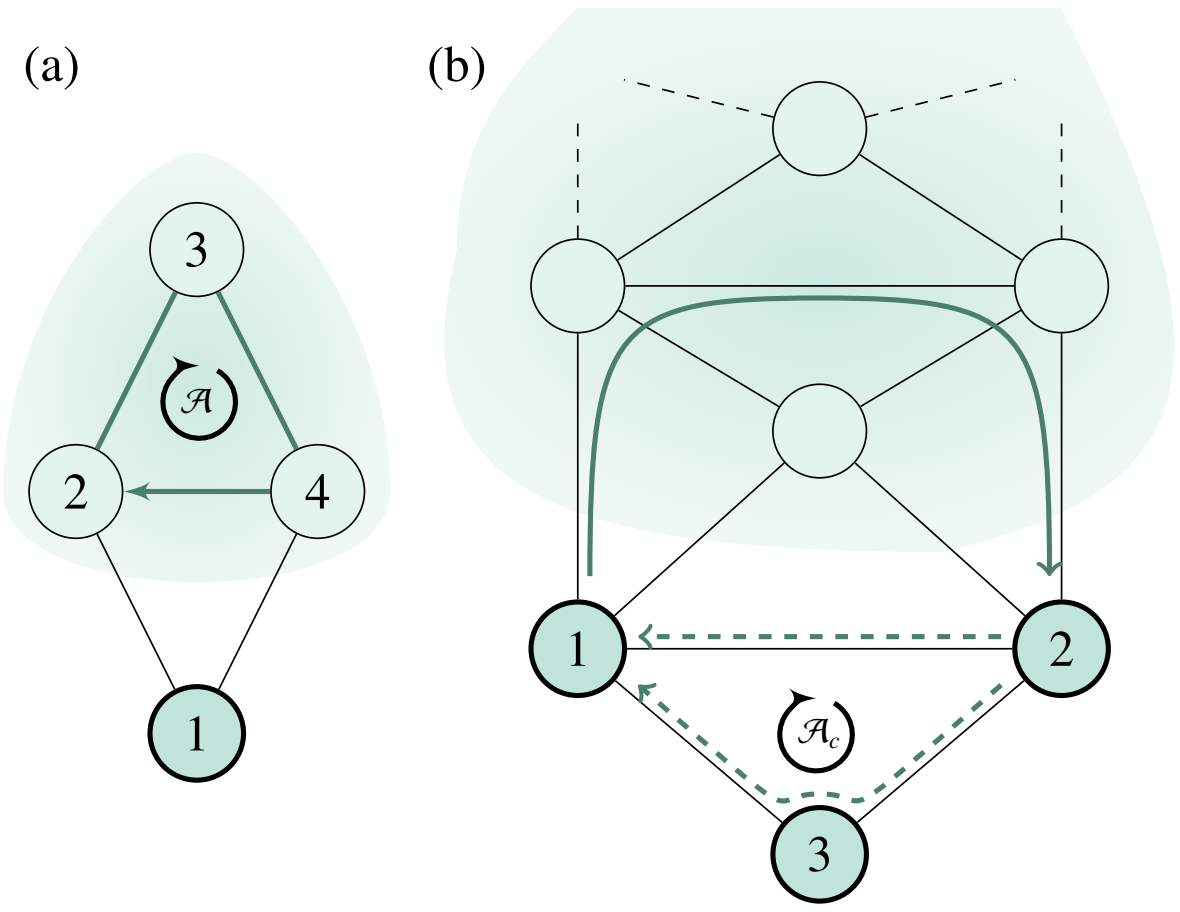

    \centering
    \includestandalone[scale=0.8]{networks_endmatter}
    \caption{Illustrative examples highlighting the necessity of additional conditions.  
    (a) An example in which neither Condition 1 nor Condition 2 is satisfied. If transition rates to and from the visible region $A = \{ 1 \}$ are much smaller than the transition rates connecting the states within $B = \{2, 3, 4\}$, a single excursion can complete the cycle $(2342)$ multiple times. Counting observables defined entirely within $B$, e.g., $Q_{23} = \hat{N}_{23} - \hat{N}_{32}$, can in principle attain arbitrarily high values, corresponding to excursions that complete the cycle multiple times. In such cases, $\ln [P(q)/P(-q)]$ can exceed the affinity $\mathcal{A}$, which only quantifies the entropy production of a single cycle.
    (b) Explanation why detailed balance of region $A$ is a necessary assumption in the case $|A| > 1$: In the illustrated network, region $A$ comprises states $1,2,3$. An excursion from state $1$ to state $2$ can be completed to a cycle through $A$ along either $2 \to 1$ or $2 \to 3 \to 1$.  The two completions have the same affinity when the internal cycle $1\to2\to3\to1$ has zero affinity.}
    \label{fig:fig_endmatter}
\end{figure}

\textit{Affinity bounds: Intuitive reason for Conditions 1 and 2}---The additional requirement of Conditions 1 or 2 to establish the affinity bound \eqref{eq:bound-max-affinity} might seem ad hoc, but to see that they cannot be relaxed, we sketch an example in Figure \ref{fig:fig_endmatter}\,a). Intuitively, repeatedly completing cycles within $B$ that contribute to the counting observable can cause $\ln [P(q)/P(-q)]$ to exceed the affinity of a single cycle. This phenomenon can be avoided if one restricts either the topology of B by excluding the existence of cycles completely within $B$, or the counting observables by limiting their increments to boundary edges between $A$ and $B$. These amendments give rise to Conditions 1 and 2, respectively. 

\textit{Derivation from Condition 1 for $|A| = 1$}--- Starting from Condition 1, we note that any excursion $\mathcal{E}$ with $\hat{Q}_\mathcal{E} \neq 0$ traverses exactly one cycle $c(\mathcal{E})$ (notice that several distinct cycles may share the same observable outcome $q$), plus some wanderings that do not contribute to $\hat{Q}_\mathcal{E}$, affinity or entropy production of the excursion. The entropy production of this excursion is then equal to its affinity, $\Delta s[\mathcal{E}] = \mathcal{A}_{c(\mathcal{E})}$.
In Ref.~\cite{Fiusa2025}, a fluctuation theorem (FT) at the excursion level was proven, where $P[\mathcal{E}] / P[\tilde{\mathcal{E}}] = \exp \Delta s[\mathcal{E}]$. Moreover, time reversal defines a bijective map onto excursions, and define $\mathcal{A}_\text{max}(q) = \max_{c : \hat{Q}(c) = q} \mathcal{A}_c$ the maximal affinity of cycles producing $q$. Plugging all together, it follows
\begin{align}
    P(-q) &= \sum_{\mathcal{E} : \hat{Q}_\mathcal{E} =q} P[\tilde{\mathcal{E}}] = \sum_{\mathcal{E} : \hat{Q}_\mathcal{E} =q} e^{-\Delta s[\mathcal{E}]} P[\mathcal{E}] = \sum_{\mathcal{E} : \hat{Q}_\mathcal{E} =q} e^{-\mathcal{A}_{c(\mathcal{E})}} P[\mathcal{E}] \nonumber\\ &\geq e^{-\mathcal{A}_\text{max} (q)} \sum_{\mathcal{E} : \hat{Q}_\mathcal{E} =q}  P[\mathcal{E}] = e^{-\mathcal{A}_\text{max} (q)} P(q).
    \label{eq:endmatter:aux1}
\end{align}
Rearranging then provides the lower bound
\begin{equation}
\label{eq:end-matter-bound-max-affinity}
    \ln \frac{P(q)}{P(-q)} \leq \mathcal{A}_\text{max} (q) \leq \mathcal{A}_\text{max}.
\end{equation}
Note that despite being the same result, Eqs.~\eqref{eq:bound-max-affinity} and~\eqref{eq:end-matter-bound-max-affinity} have different assumptions. In the former, we request that the counting observable $\hat{Q}$ adds counts only in the border between $A$ and $B$, without any assumptions about the topology of $B$; whereas in the latter, there are no assumptions over the counting observable (other than that being time-antisymmetric) but we now assume that region $B$ has no closed hidden cycles. In \cite{supp}, we prove, using spanning trees and graph theoretic arguments, that Eq.~\eqref{eq:bound-max-affinity} holds for arbitrary topology of $B$.

\textit{The case $|A| > 1$}---
If $A$ contains more than one state, it is not immediately clear what the natural generalization of the result \eqref{eq:end-matter-bound-max-affinity} is, because excursions need not be closed loops in this case. 

One way to generalize the result is to replace the cycle affinity with the quantity $\Delta s[\mathcal{E}]$, but the resulting inequality is a lower bound on the entropy production of paths with certain properties, similar to results established in Refs. \cite{JannaPRR2024, Maier2025}. For example, Eq. \eqref{eq:endmatter:aux1} gives
\begin{align}
    P(-q) \geq  P(q) \min_{\mathcal{E}: \hat{Q}_\mathcal{E} =q} e^{- \Delta s[\mathcal{E}]}
\end{align}
where the minimum is finite because Condition 1 ensures that an excursion does not contain any cycles.

A different way to generalize Eq. \eqref{eq:end-matter-bound-max-affinity} is to identify necessary assumptions under which the concept of an affinity retains its meaning. A reasonable necessary assumption is that $A$ is connected and at detailed balance in the sense that $\mathcal{A}_c = 0$ for every cycle $c$ fully contained in $A$. The second statement is equivalent to the existence of a function $r_x$ such that $r_x W_{x'x}=r_{x'} W_{xx'}$ is satisfied for $x,x'\in A$. If $A$ does not satisfy this condition, an excursion itself does not give rise to a well-defined cycle affinity: If an excursion without self-crossings is extended into a cycle through a path in $A$, the affinity of the resulting cycle generally depends on the choice of the path. The condition of detailed balance ensures that one can attribute a well-defined cycle affinity to an excursion that is independent of the chosen path. An intuitive illustration is provided in Figure \ref{fig:fig_endmatter}\,b). 

Under these conditions, Eq. \eqref{eq:end-matter-bound-max-affinity} can be generalized to the case $|A| > 1$, but acquires a correction term. This term is operationally accessible if region $A$ can be accessed. The result and its derivation from Condition 2 is found in \cite{supp}.

\widetext

\newpage 

\begin{center}
\vskip0.5cm
{\large\bfseries Supplemental Material for\\[2pt] ``Thermodynamic inference from excursions into hidden states: theory and quantum-dot experiment''}
\vskip0.4cm
Guilherme Fiusa,$^{1,2}$ Jann van der Meer,$^{3}$ Tan Van Vu,$^{4}$ Vivek Wadhia,$^{5}$\\ Natalia Ares,$^{5}$ Gabriel T. Landi,$^{1,2}$ and Pedro E. Harunari$^{6}$
\vskip0.2cm
{\small\itshape
$^{1}$Department of Physics and Astronomy, University of Rochester, Rochester, New York 14627, USA\\
$^{2}$University of Rochester Center for Coherence and Quantum Science, Rochester, New York 14627, USA\\
$^{3}$Department of Physics No. 1, Graduate School of Science, Kyoto University, Kyoto 606-8502, Japan\\
$^{4}$Center for Gravitational Physics and Quantum Information,
Yukawa Institute for
Theoretical Physics,\\ Kyoto University, Kitashirakawa Oiwakecho, Sakyo-ku, Kyoto 606-8502, Japan\\
$^{5}$Department of Engineering Science, University of Oxford, Parks Road, Oxford OX1 3PJ, United Kingdom\\
$^{6}$Aix Marseille Universit\'e, CNRS, CINAM, Turing Center for Living Systems, 13288 Marseille, France\par}
\vskip0.2cm
{\small (Dated: \today)}
\end{center}

\vskip0.4cm

\setcounter{section}{0}
\setcounter{equation}{0}
\setcounter{figure}{0}
\setcounter{table}{0}
\setcounter{page}{1}
\renewcommand{\thefigure}{S\arabic{figure}}
\counterwithout{equation}{section}
\renewcommand{\theequation}{S\arabic{equation}}
\setcounter{secnumdepth}{3}
\renewcommand{\thesection}{\arabic{section}}
\renewcommand{\thesubsection}{\Alph{subsection}}

\section{Proof of detailed balance inference}
\label{appendix:proof-db}
An excursion is a path $\mathcal{E} = (x_1, t_1, x_2, t_2, \dots, x_{N-1}, t_{N-1}, x_N)$, with $x_1, x_N \in A$ and $x_i \in B$ for $1 < i < N$, the system residing for a time $t_i$ in state $x_i$. The residence times are random variables, but in the Appendices we omit the hats for residence times. Since the excursion begins at the jump $x_1 \to x_2$ and ends at the jump $x_{N-1} \to x_N$, its weight collects the $N-1$ jump factors and the residence times of the interior states only, i.e.
\begin{equation}
  P(\mathcal{E})
  = W_{x_N x_{N-1}} e^{-\Gamma_{x_{N-1}} t_{N-1}} \cdots
    W_{x_3 x_2} e^{-\Gamma_{x_2} t_2} W_{x_2 x_1},
  \qquad
  \Gamma_x = \sum_{y (\neq x)} W_{yx},
  \label{eq:A-weight}
\end{equation}
and its duration is the time spent in $B$, which then reads $\hat{T}[\mathcal{E}] = \sum_{i=2}^{N-1} t_i$. Note that the residence times $t_1$ and $t_N$ in the visible
boundary states are not part of the excursion. The counting observable
evaluated on $\mathcal{E}$ is $\hat{Q}[\mathcal{E}] = \sum_{x,y} \nu_{xy} \hat N_{xy}[\mathcal{E}]$, with
$\hat N_{xy}[\mathcal{E}]$ the number of jumps $y \to x$ in the excursion $\mathcal{E}$, the two boundary jumps included.

Time reversal acts on excursions in the following way
\begin{equation}
  \tilde{\mathcal{E}} = (x_N, t_{N-1}, x_{N-1}, \dots, t_2, x_2, x_1),
  \label{eq:A-reversal}
\end{equation}
that is, it reverses the order in which the states are visited while leaving the residence times attached to each interior state unchanged. Because the interior states are the same set, $\tilde{ \mathcal{E}}$ is again an excursion, now running from $x_N$ to $x_1$, with the same duration $\hat{T}[\tilde{\mathcal{E}}] = \hat{T}[\mathcal{E}]$ and the same residence
factors as $\mathcal{E}$. Its weight is given by
\begin{equation}
  P(\tilde{\mathcal{E}})
  = W_{x_1 x_2} e^{-\Gamma_{x_2} t_2} W_{x_2 x_3} \cdots
   e^{-\Gamma_{x_{N-1}} t_{N-1}} W_{x_{N-1} x_N}.
  \label{eq:A-weight-rev}
\end{equation}
Reversal is an involution, which means it is a bijection of the set of excursions onto itself, and it maps the excursions running $x_1 \to x_N$ bijectively onto those running $x_N \to x_1$. For time-antisymmetric weights, $\nu_{xy} = -\nu_{yx}$, one has $\hat N_{xy}[\tilde{\mathcal{E}}] = \hat N_{yx}[\mathcal{E}]$ and therefore the counting observables have the property
\begin{equation}
  \hat{Q}[\tilde{\mathcal{E}}] = -\,\hat{Q}[\mathcal{E}].
  \label{eq:A-Qflip}
\end{equation}
Dividing Eq.~\eqref{eq:A-weight} by Eq.~\eqref{eq:A-weight-rev}, every residence time factor cancels and only the jump rates survive, yielding
\begin{equation}
  \ln \frac{P(\mathcal{E})}{P(\tilde{\mathcal{E}})}
  = \sum_{i=1}^{N-1} \ln \frac{W_{x_{i+1} x_i}}{W_{x_i x_{i+1}}}
  = \Delta s[\mathcal{E}],
  \label{eq:A-pathratio}
\end{equation}
which is precisely the heat dissipated along the excursion, see Eq.~\eqref{eq:dissipation-one-excursion}. 
Suppose now that the dynamics satisfies detailed balance, $W_{xy}\pi_y = W_{yx}\pi_x$ for every pair of states, which is the hypothesis whose violation we wish to detect. Each ratio in Eq.~\eqref{eq:A-pathratio} can then be rewritten as a ratio of stationary occupations, following $W_{x_{i+1}x_i}/W_{x_i x_{i+1}} = \pi_{x_{i+1}}/\pi_{x_i}$. Note that all intermediate states cancel and only the two endpoints survive, so we obtain
\begin{equation}
  \Delta s[\mathcal{E}] = \ln \frac{\pi_{x_N}}{\pi_{x_1}},
  \qquad \text{so that} \qquad
  P(\mathcal{E}) = \frac{\pi_{x_N}}{\pi_{x_1}} \, P(\tilde{\mathcal{E}}).
  \label{eq:detailed-balance-stationary}
\end{equation}
Note that the essential point is not the particular value of this factor but rather its support. 
Under detailed balance the path ratio depends on the excursion only through the pair of visible states at which it starts and ends, and is blind to everything that happened in $B$. Equivalently, since $\ln(\pi_{x_N} /
\pi_{x_1})$ is minus the change in stochastic system entropy along the excursion, Eq.~\eqref{eq:detailed-balance-stationary} states that every excursion produces zero total entropy pathwise.

Because there is still dependence on boundary states in Eq.~\eqref{eq:detailed-balance-stationary}, we resolve the excursion statistics with respect to them. For $x, y \in A$, let
\begin{equation}
  P_{xy}(q,t)
  := \frac{1}{K_{xy}} \sum_{\mathcal{E} : x \to y}
    D[\mathcal{E}] \, P(\mathcal{E}) \,
    \delta_{q, \hat{Q}[\mathcal{E}]} \, \delta\!\left( t - \hat{T}[\mathcal{E}] \right)
  \label{eq:A-joint}
\end{equation}
be the joint distribution of the counting observable and the duration over excursions that leave $A$ from $x$ and return to $A$ at $y$, with $K_{xy}$ the normalization, and $\sum_{\mathcal{E}}D[\mathcal{E}]$ is the shorthand for the sum of all excursions: it runs over the number of jumps, over the sequence of states visited in region $B$, and over all residence times spent within states in $B$. Importantly, it is invariant under time-reversal $D[\mathcal{E}] = D[\tilde{\mathcal{E}}]$. Substituting Eq.~\eqref{eq:detailed-balance-stationary} into Eq.~\eqref{eq:A-joint}, the boundary states are
held fixed by the conditioning, so the factor $\pi_y/\pi_x$ is constant
throughout the sum and can be taken out:
\begin{align}
P_{xy}(q,t)
&= \frac{\pi_y/\pi_x}{K_{xy}} \sum_{\mathcal{E}:x\to y} D[\mathcal{E}]\, P(\tilde{\mathcal{E}})\,
\delta_{q,\hat{Q}[\mathcal{E}]} \delta\!\left(t - \hat{T}[\mathcal{E}]\right) \nonumber \\
&= \frac{\pi_y/\pi_x}{K_{xy}} \sum_{\tilde{\mathcal{E}}:y\to x} D[\tilde{\mathcal{E}}]\, P(\tilde{\mathcal{E}})\,
\delta_{-q,\hat{Q}[\tilde{\mathcal{E}}]} \delta\!\left(t - \hat{T}[\tilde{\mathcal{E}}]\right) \nonumber \\
&= \frac{\pi_y}{\pi_x}\, \frac{K_{yx}}{K_{xy}}\, P_{yx}(-q,t),
  \label{eq:A-joint-relation}
\end{align}
where we used the bijection $\mathcal{E} \mapsto \tilde{\mathcal{E}}$ between excursions $x \to y$ and $y \to x$ to relabel the second line, together with $\hat{Q}[\tilde{\mathcal{E}}] = -\hat{Q}[\mathcal{E}]$ and $\hat{T}[\tilde{\mathcal{E}}] = \hat{T}[\mathcal{E}]$ from Eqs.~\eqref{eq:A-Qflip} and \eqref{eq:A-reversal}.

Marginalizing Eq.~\eqref{eq:A-joint-relation} over the duration gives $P_{xy}(q) = (\pi_y/\pi_x)(K_{yx}/K_{xy}) P_{yx}(-q)$, and the two prefactors are the same in both relations. Expanding the conditional distributions $P_{xy}(t \mid q) = P_{xy}(q,t)/P_{xy}(q)$, we obtain 
\begin{equation}
\frac{P_{xy}(t \mid q)}{P_{yx}(t \mid -q)}
= \frac{P_{xy}(q,t)\, P_{yx}(-q)}{P_{yx}(-q,t)\, P_{xy}(q)}
= \frac{(\pi_y/\pi_x)(K_{yx}/K_{xy}) P_{yx}(-q,t)\, P_{yx}(-q)}
{P_{yx}(-q,t)\, (\pi_y/\pi_x)(K_{yx}/K_{xy}) P_{yx}(-q)} = 1.
  \label{eq:A-conditional}
\end{equation}
Note that the ratio of stationary occupations at the endpoints is the only trace
left of the dynamics in $B$ after Eq.~\eqref{eq:detailed-balance-stationary}. Therefore, it is divided out by the conditioning, and detailed balance manifests itself as an exact symmetry of the duration statistics at fixed magnitude of the counting observable.

If region $A$ has a single state, which is the setting of the double quantum dot we discussed in this Letter, the result then simplifies. In this case, every excursion starts and ends at the same state, so $\pi_y/\pi_x = 1$ and $K_{xy} = K_{yx}$ identically, and Eq.~\eqref{eq:A-conditional}
becomes a statement about a symmetry of the joint distribution itself,
\begin{equation}
  P(q,t) = P(-q,t),
  \qquad\text{hence}\qquad
  P(q) = P(-q)
  \qquad\text{and}\qquad
  \frac{P(t \mid q)}{P(t \mid -q)} = 1,
  \label{eq:A-unitA}
\end{equation}
the last of which is Eq.~\eqref{eq:DB-symmetry} of the main text. 
Detailed balance therefore constrains the frequencies of the outcomes and the duration
statistics at fixed outcome separately, and the two can be tested independently on the same data.
Note that the first equality in Eq.~\eqref{eq:A-unitA} is the equilibrium limit of the affinity bound,
Eq.~\eqref{eq:bound-max-affinity}, since all cycle affinities vanish under detailed balance.
Moreover, we emphasize that the derivation holds for any time-antisymmetric $\hat Q$, since
Eq.~\eqref{eq:A-Qflip} is the only property of the observable that was used.
If the dynamics additionally obeys local detailed balance with respect to a single reservoir at inverse temperature $\beta$, the stationary state is a Gibbs state and $\pi_y/\pi_x = e^{-\beta \Delta E}$ with $\Delta E = E_y - E_x$. Nothing in the argument above relies on this.

\section{Proof of the affinity bound}

The first section contains the proof of Eq.~\eqref{eq:bound-max-affinity} under Condition 2 for $|A|=1$, which applies to the DQD experiment. After discussing a formulation in terms of fundamental cycles, we explain the necessary amendments for the $|A| > 1$ case.

\subsection{Derivation of Eq. \eqref{eq:bound-max-affinity} from Condition 2 for $|A|=1$}

Assuming $A=\{x\}$, all excursions begin and end at the same state $x \in A$. 
Since the result concerns only counts of the counting observables, the times that appear within the excursion are not relevant and for simplicity we omit those here. Moreover, let us consider the first and last visited states in $B$ to be denoted as $y_1$ and $y_2$; an excursion therefore can be written as $\mathcal{E}= (x, y_1, \dots, y_2, x)$.
Condition 2 assumes that a counting observable only has increments at the boundary between $A$ and $B$, $\nu_{xy} = 0$ when $x,y \in B$. Thus, the counting observable of this excursion only depends on $y_1$ and $y_2$, $\hat{Q}_{\mathcal E} = \nu_{y_1 ,x} + \nu_{x, y_2} := \hat{Q}(y_1, y_2)$. The probability $P(y_1,y_2) $ of an excursion with these two endpoints can be evaluated through polynomials of rooted trees that account for all paths within $B$ leading to such an excursion. If $\mathbb{T}$ is a rooted spanning tree (a sub-network connecting all vertices, with no cycles, and all edges point towards the root), $\tau (\mathbb{T})$ is the product of all its transition rates. Let $\mathbb{T}^x$ be the set of all possible spanning trees rooted at $x$,  and let $\mathbb{T}^x_{y_1\to y_2}\subseteq \mathbb{T}^x$ contain those whose branch from $y_1$ reaches $x$ from $y_2$. These quantities are useful in diagrammatic approaches to evaluate propagation probabilities, fluxes and first-passage times~\cite{Hill1966}. By the harmonic tree formula [Eq.~(1.9) of Ref.~\cite{Pitman2018}], which follows from Kirchhoff's matrix forest theorem (Theorem~1.2 therein), applied to the chain in which every jump $y \to x$ is redirected to a separate absorbing copy of $x$, so that each forest rooted at the copies becomes a tree in $\mathbb{T}^x$,
\begin{equation}
    P(y_1,y_2) = P(y_1) P(y_2|y_1) = \frac{W_{y_1 x}}{\Gamma_x} \frac{\sum_{\mathbb{T} \in \mathbb{T}^x_{y_1 \to y_2}} \tau (\mathbb{T})}{\sum_{\mathbb{T} \in \mathbb{T}^x} \tau (\mathbb{T})} := \frac{W_{y_1 x}}{\Gamma_x} \frac{ \omega^x_{y_1 \to y_2}}{\omega^x},
    \label{eq:affinity-supp:aux1}
\end{equation}
where $P(y_1)$ and $P(y_2|y_1)$ denotes the probability that an excursion leaving $x$ starts at $y_1$ and an excursion starting in $y_1$ leaves region $B$ through state $y_2$, respectively.
The probability for a counting observable outcome $q$ is $P(q) = \sum_{y_1,y_2} P(y_1,y_2) \delta_{\hat{Q}(y_1, y_2), q}$. Every tree $\mathbb{T} \in \mathbb{T}^x_{y_1 \to y_2}$ singles out one simple cycle $c_\mathbb{T}$ through $x$: its branch from $y_1$ to $y_2$ and closed by $x \to y_1$; and $\hat{Q}(c_\mathbb{T}) = \hat{Q}(y_1,y_2)$. Reversing that branch while keeping every other edge of $\mathbb{T}$ untouched yields $\bar{\mathbb{T}}$, forming the bijective map $\mathbb{T}^x_{y_1 \to y_2} \leftrightarrow \mathbb{T}^x_{y_2 \to y_1}$, and only the rates along $c_\mathbb{T}$ are affected, so $W_{y_2 x} \tau(\bar{\mathbb{T}}) = W_{y_1 x} \tau(\mathbb{T}) e^{-\mathcal{A}_{c_\mathbb{T}}}$, with $\mathcal{A}_{c_\mathbb{T}}$ the affinity of $c_\mathbb{T}$.

By evaluating the probability of $-q$ and applying this reversal map,
\begin{align}
    P(-q) &= \sum_{y_1,y_2} \delta_{\hat{Q}(y_2,y_1),-q} \frac{W_{y_2 x}}{\Gamma_x} \frac{\sum_{\mathbb{T} \in \mathbb{T}^x_{y_2 \to y_1}} \tau (\mathbb{T})}{\omega^x} \nonumber\\
    &= \sum_{y_1,y_2} \delta_{\hat{Q}(y_1,y_2),q} \frac{W_{y_1 x}}{\Gamma_x} \frac{\sum_{\mathbb{T} \in \mathbb{T}^x_{y_1 \to y_2}} e^{-\mathcal{A}_{c_\mathbb{T}}} \tau (\mathbb{T})}{\omega^x} \geq e^{-\mathcal{A}_\text{max}(q)} P(q),
\end{align}
where $\mathcal{A}_\text{max}(q)$ is the largest affinity among the simple cycles through $x$ with counting observable $q$, and $\mathcal{A}_\text{max}$ is the largest for any $q$. Therefore, we have the bound 
\begin{equation}
    \ln \frac{P(q)}{P(-q)} \leq \mathcal{A}_\text{max} (q) \leq \mathcal{A}_\text{max}
    \label{eq:app-aff:aux1}
,\end{equation}
stated as Eq.~\eqref{eq:bound-max-affinity} in the main text. It is worth noting that cycles lying entirely within $B$ never appear among the $c_\mathbb{T}$ and are invisible to the argument. 

\subsection{Restating the bound for fundamental cycles}

The same bound can be phrased in the Schnakenberg basis~\cite{schnakenberg1976}. Fix a spanning tree $\mathcal{T}$ of the network, let $\alpha$ label its $n_\text{f}$ chords and $\mathcal{A}_\alpha$ the affinities of the associated fundamental cycles, so that entropy production can be decomposed as $\sigma = \sum_\alpha j_\alpha \mathcal{A}_\alpha$, and let $\mathcal{A}^\text{f}_\text{max} :=\max_\alpha |\mathcal{A}_\alpha|$. Any simple cycle decomposes as $\mathcal{A}_c = \sum_\alpha n_\alpha \mathcal{A}_\alpha$, with $n_\alpha \in \{ 0, \pm 1\}$. Since any simple-cycle affinity is a combination of at most $n_\text{f}$ fundamental cycles, Eq. \eqref{eq:app-aff:aux1} implies
\begin{equation}
    \frac{1}{n_\text{f}}\ln \frac{P(q)}{P(-q)} \leq \mathcal{A}^\text{f}_\text{max}
\end{equation}
for any fundamental cycle decomposition. If there are no hidden cycles in $B$ (Condition 1 in the main text), there is a particular fundamental cycle decomposition for which this result can be improved. Since $B$ has no hidden cycles, the collection of edges connecting states within $B$ can be taken as part of a spanning tree. The chords are the network edges not already contained in this spanning tree, so the only chords a cycle containing $x$ can pass through are edges that connect the state $x$ defining $A$ to region $B$. A cycle through $x$ traverses exactly two such edges (one leaving and one entering $x$), so it contains at most two chords, therefore the previous result simplifies to $\left(1/2\right) \ln [P(q) / P(-q)] \leq \mathcal{A}^\text{f}_\text{max}$ for this particular fundamental cycle decomposition.

\subsection{Operationally accessible correction for the case $|A| > 1$}

The reasoning via spanning trees can be extended to the case that region $A$ contains more than one state, but as explained in the End Matter the interpretation as a bound on cycle affinities is only unambiguous if $A$ is in detailed balance.
We assume that $A$ is connected and that the transition rates within $A$ are at detailed balance in the sense that there one can find a function $r_x$ such that $r_xW_{x'x}=r_{x'}W_{xx'}$ is satisfied for $x,x'\in A$.
Note that in general $r_x$ will differ from $\pi_x$, the stationary distribution of the entire network.
We quantify the difference by defining
\begin{equation}
    h_x:=\ln\frac{\pi_x}{r_x}
    \label{eq:app-aff:def-h}
.\end{equation}
As discussed in the End Matter, the existence of $r_x$ in $A$ allows completing an excursion by appending a path in $A$. For example, a microscopic trajectory $\gamma$ that gets coarse-grained into an excursion $x\to x'$ gives rise to an affinity
$\mathcal A_\gamma=
\Delta s[\gamma]+\ln(r_x/r_{x'})$
that does not depend on the path from $x'$ to $x$ through $A$.

We now discuss the affinity bound, focusing only on the necessary modification to the previous argument. If $A$ has more than one state, a counting observable $\hat{Q}_{\mathcal E} = \hat{Q}(x_1, y_1, y_2, x_2)$ evaluated on an excursion $\mathcal E = (x_1, y_1, ..., y_2, x_2)$ depends on the entry and exit states within $B$, $y_1, y_2$, as well as the states $x_1$ and $x_2$ through which region $A$ was exited and entered, respectively. The probability that an excursion that leaves $A$ through $x_1$ is of the form $\mathcal E = (x_1, y_1, ..., y_2, x_2)$ is given by
\begin{align*}
    P(y_1,y_2,x_2|x_1) & = P(y_1|x_1)P(y_2|y_1) P(x_2|y_2) \\
    & = \frac{W_{y_1 x_1}}{\sum_{y_i \in B} W_{y_ix_1}} \frac{\sum_{\mathbb{T} \in \mathbb{T}^{x^*}_{y_1 \to y_2}} \tau (\mathbb{T})}{\sum_{\mathbb{T} \in \mathbb{T}^{x^*}} \tau (\mathbb{T})} \frac{W_{x_2 y_2}}{\sum_{x_i \in A} W_{x_iy_2}}
.\end{align*}
Apart from the additional dependence on the final state of the excursion $x_2$, there is an important modification compared to Eq. \eqref{eq:affinity-supp:aux1}: The rooted spanning trees are defined on an auxiliary network that contains region $B$ but replaces the entire region $A$ with an absorbing state, i.e., any transition $y \to x$ gets redirected into a transition $y \to x^*$ into a single auxiliary state $x^*$ with transition rate $\sum_{x\in A}W_{xy} = W_{{x^*}y} =: \kappa_y$. Intuitively, the dynamics from $y_1$ to $y_2$ within $B$ is not affected by redirecting outgoing transition rates into state $A$, as long as they still terminate the dynamics within $B$.

We define $\Gamma_x^B:=\sum_{y\in B}W_{yx}$ and $K:=\sum_{x\in A}\pi_x\Gamma_x^B$ to express the stationary probability of excursions of the form $\mathcal E = (x_1, y_1, ..., y_2, x_2)$ as
\begin{equation}
\begin{aligned}
P_{x_1x_2}(y_1,y_2)
&=\frac{\pi_{x_1}\Gamma_{x_1}^B}{K}
P(y_1,y_2,x_2|x_1)\\
&=\frac{\pi_{x_1}W_{y_1x_1}}{K}
\frac{\omega^{x^*}_{y_1\to y_2}}{\omega^{x^*}}
\frac{W_{x_2y_2}}{\kappa_{y_2}},
\end{aligned}
\end{equation}
where $\omega^{x^*}$ and $\omega^{x^*}_{y_1\to y_2}$ are defined as in the single-state case, but with root $x^*$.

We now make use of the function $r_x$ defined in Eq. \eqref{eq:app-aff:def-h} and the condition that $A$ is at detailed balance. Each tree $\mathbb T\in\mathbb T^{x^*}_{y_1\to y_2}$ defines a simple cycle $c_{\mathbb T}$ by completing the branch $x_1\to y_1\to\cdots\to y_2\to x_2$
with a path from $x_2$ to $x_1$ within $A$. Its affinity is independent of the closing path, and its counting outcome $\hat Q(x_1,y_1,y_2,x_2)$ remains well-defined because the counting observable is insensitive to edges within $A$. 

As in the single-state proof, reversing the branch gives $\bar{\mathbb T}\in\mathbb T^{x^*}_{y_2\to y_1}$, but unlike in the single-state case we introduce a correction term $h_x:=\ln(\pi_x/r_x)$. The corresponding tree weights satisfy
\begin{equation}
\begin{aligned}
&\frac{\pi_{x_2}W_{y_2x_2}W_{x_1y_1}}{\kappa_{y_1}}
\tau(\bar{\mathbb T}) =
\frac{\pi_{x_1}W_{y_1x_1}W_{x_2y_2}}{\kappa_{y_2}}
\tau(\mathbb T)
e^{-\mathcal A_{c_{\mathbb T}}-h_{x_1}+h_{x_2}}.
\end{aligned}
\end{equation}
The factors $\kappa_{y_1},\kappa_{y_2}$ cancel against corresponding terms contained in the tree weights, because trees rooted at $x^*$ contain a factor $\kappa_y$ from the edge $y \to x^*$. Compared with the single-state case, the
additional factor $e^{-h_{x_1}+h_{x_2}}$ remains.

We can define a discrepancy term $\delta_A:=\max_{x\in A}h_x-\min_{x\in A}h_x$ and apply the same summation and partial-reversal argument as in the case that $|A| = 1$ to obtain
\begin{equation}
P(-q) \geq e^{-\mathcal A_{\max}(q)-\delta_A}P(q),
\end{equation}
and therefore
\begin{equation}
\ln\frac{P(q)}{P(-q)}-\delta_A \leq\mathcal A_{\max}(q) \leq\mathcal A_{\max}.
\end{equation}
We note that the correction vanishes only when $\pi_x\propto r_x$ in $A$, in particular when detailed balance holds on all internal edges of $A$ in the stationary state. It is also important to note that $h_x$ and $\delta_A$ are operationally accessible quantities if $A$ is accessible. The stationary distribution $\pi_x$ and internal transition rates $W_{x_2x_1}$ can be determined from a long, stationary trajectory in the full network. Solving the linear equations $r_xW_{x'x}=r_{x'}W_{xx'}$ then determines $r_x$, which in turns allows calculating $h_x$ from Eq. \eqref{eq:app-aff:def-h}. 

\section{Proof of entropy production bound for general cases $|A|\ge 1$}
\label{appendix:proof-bound-ep}
Let $\gamma_t=(x_1,t_1,x_2,t_2,\dots)$ be a typical path that is generated by the Markov jump process for a long time $t$, where $x_1\in A$.
Here, the system stays at state $x_i$ for a duration $t_i$, then jumps to the next state $x_{i+1}$.
In general, $\gamma_t$ can be decomposed into a sequence of independent complete excursions, interspersed with paths consisting of states belonging to $A$, followed by an incomplete residual path at the end; that is, $\gamma_t=(\mathcal{E}_1,F_1,\mathcal{E}_2,F_2,\dots,\mathcal{E}_{N_t},F_{N_t},\mathcal{I})$.
Let $\Delta s_k$ and $\Delta s_k'$ denote heat dissipation along the excursion $\mathcal{E}_k$ and path $F_k$, respectively, and $\Delta s_{\rm res}$ be heat dissipation along the residual path $\mathcal{I}$.
Note that it is possible that $F_{k}$ is empty for some $k$; in this case, $\Delta s_{k}'=0$.
Specifically, if $\mathcal{E}_k=(x_1,t_1,x_2,t_2,\dots,x_{N-1},t_{N-1},x_N)$, where $x_1,x_N\in A$ and $x_i\in B$ for $1<i<N$, then
\begin{equation}\label{eq:dissipation-one-excursion}
    \Delta s_k=\sum_{i=1}^{N-1}\ln\frac{W_{x_{i+1}x_i}}{W_{x_ix_{i+1}}}.
\end{equation}
Additionally, if $F_k=(x_1,t_1,x_2,t_2,\dots,x_{N-1},t_{N-1},x_N)$, where $x_i\in A$ for all $1\le i\le N$, then
\begin{equation}
    \Delta s_k'=\sum_{i=1}^{N-1}\ln\frac{W_{x_{i+1}x_i}}{W_{x_ix_{i+1}}}.
\end{equation}

Due to the ergodicity of Markov jump processes, the entropy production rate can be
expressed as
\begin{equation}  \sigma=\lim_{t\to\infty}\dfrac{1}{t}\ln\frac{P(\gamma_t)}{P(\tilde{\gamma}_t)},
\end{equation}
where $P(\gamma_t)=e^{-\Gamma_{x_1}t_1}W_{x_2x_1}e^{-\Gamma_{x_2}t_2}W_{x_3x_2}\dots$ and $\Gamma_{x} = \sum_{y(\neq x)}W_{yx}$ is the escape rate from $x$.
Since $N_t\to\infty$ as $t\to\infty$, the entropy production rate can be calculated as follows:
\begin{align}\label{eq:app:proof-bound-ep:aux3}
    \sigma&=\lim_{t\to\infty}\frac{\Delta s_{\rm res} + \sum_{k=1}^{N_t}\Delta s_k}{t} + \lim_{t\to\infty}\frac{\sum_{k=1}^{N_t}\Delta s_k'}{t}\notag\\
    &=\lim_{t\to\infty}\qty(\frac{\Delta s_{\rm res} + \sum_{k=1}^{N_t}\Delta s_k}{N_t})\qty(\frac{N_t}{t}) + \sum_{x,y\in A}W_{xy}\pi_y\ln\frac{W_{xy}\pi_y}{W_{yx}\pi_x}\notag\\
    &=K{\ev{\Delta s_k}} + \sigma_A.
\end{align}
Here, $\sigma_A$ denotes the partial entropy production rate associated with region $A$.
In the case where $|A|=1$ or there are no direct jumps between states in $A$ even if $|A|>1$, $\sigma_A=0$, and the same result for the $|A|=1$ case is recovered.

For each excursion $\mathcal{E}=(x_1,t_1,\dots,x_{N-1},t_{N-1},x_N)$, its path weight $P(\mathcal{E})$ is given by Eq.~\eqref{eq:A-weight}.
Since the probability that the system starts from $A$ and returns to $A$ at some time is equal to $1$, the average heat dissipation of each excursion can be evaluated as
\begin{align}
    \ev{\Delta s_k}&=\sum_{\mathcal{E}}D[\mathcal{E}]P(\mathcal{E})\ln\frac{P(\mathcal{E})}{P(\tilde{\mathcal{E}})}\ge\sum_{q}P(q)\ln\frac{P(q)}{P(-q)},
    \label{eq:app:proof-bound-ep:aux1}
\end{align}
where we applied the log-sum inequality to obtain the last inequality.
Consequently, we obtain the following lower bound for entropy production rate in the general case $|A|\ge 1$:
\begin{equation}
    \sigma\ge K{D_{\rm KL}(P(q)\|P(-q))} + \sigma_A.
\end{equation}
Note that all the terms in the lower bound can be estimated from the trajectory data.

\section{Saturation on unicyclic networks}
\label{appendix:bound-saturation}
Let the network consist of a single cycle $c$ through all $N$ states and take $A=\{1\}$, so that $B$ is the path $2-3-\cdots-N$ and $\sigma_A=0$. Choose the counting observable supported on one boundary edge, $\nu_{1N}=-\nu_{N1}=1$ with all remaining weights vanishing, so that $\hat{Q}$ records the net number of times $c$ is completed during an excursion.
An excursion enters $B$ through $2$ or through $N$ and leaves through $2$ or through $N$. Because $B$ is a path, an excursion entering and exiting through the same end traverses every edge an equal number of times in both
orientations; it produces $\Delta s=0$ and, since its two boundary transitions cancel, $q=0$. An excursion entering through $2$ and exiting through $N$ traverses each edge of $c$ exactly once in the positive orientation, so $\Delta s=\mathcal{A}_c$ and $q=+1$, and the reverse ordering gives $\Delta s=-\mathcal{A}_c$ and $q=-1$. Note that $|q|\le1$, since completing $c$ twice would require an intermediate visit to state $1$, which terminates the excursion. Hence
\begin{equation}
  \Delta s[\mathcal{E}] = q[\mathcal{E}]\,\mathcal{A}_c ,
  \label{eq:dsq}
\end{equation}
that is, the entropy production of an excursion is a deterministic function of the counting observable alone.

Two consequences follow. First, $\ln P(\mathcal{E})/P(\tilde{\mathcal{E}})=\Delta s[\mathcal{E}]$ is constant on each level set of $q$, so the coarse graining $\mathcal{E} \mapsto q$ discards nothing and the log-sum inequality~\eqref{eq:app:proof-bound-ep:aux1} holds with equality,
\begin{equation}
  \langle\Delta s_k\rangle = \sum_q P(q)\,q\,\mathcal{A}_c .
\end{equation}
Second, time reversal maps the excursions with $q=+1$ bijectively onto those with $q=-1$ and multiplies each path weight by $e^{-\mathcal{A}_c}$, whence $P(-1)=e^{-\mathcal{A}_c}P(+1)$ and
\begin{equation}
  \ln\frac{P(q)}{P(-q)} = q\mathcal{A}_c .
  \label{eq:logratio}
\end{equation}
Substituting into Eq.~\eqref{eq:bound-ep},
\begin{equation}
  \hat\sigma_{\rm exc}
  = K\sum_q P(q)\ln\frac{P(q)}{P(-q)}
  = K\big[P(+1)-P(-1)\big]\mathcal{A}_c
  = j_c \mathcal{A}_c = \sigma ,
\end{equation}
where $j_c=K[P(+1)-P(-1)]$ is the net rate of cycle completions and the last equality is the Schnakenberg decomposition for a unicyclic network~\cite{schnakenberg1976}. The estimator therefore returns the exact dissipation rate.
The argument used only two properties. First, $B$ contains no cycles, so that any trajectory snippet within $B$ retraces edges and cancels in $\Delta s$, leaving the dissipation of an excursion fixed by its pair of boundary transitions. Second, $\hat{Q}$ resolves the orientation of the crossing, because if both boundary transitions carry equal weight then $q\equiv0$, which means that saturation is a property of the network \emph{and} the observable.

Saturation therefore extends beyond $|A|=1$. If $A$ consists of states that are consecutive along $c$, then $B$ is again a single path: every excursion either returns through the end by which it entered, contributing $\Delta s=q=0$, or crosses $B$ once, contributing $\Delta s=\pm \mathcal{A}_B$ and $q=\pm1$, where $\mathcal{A}_B$ is the affinity of the arc (i.e. trajectory snippet in a ring topology) spanned by $B$ together with its two boundary edges. Since every edge of a unicyclic network carries the same net current $j_c$, the residual term of Eq.~\eqref{eq:app:proof-bound-ep:aux3} evaluates to $\sigma_A=j_c \mathcal{A}_A$ with $\mathcal{A}_A$ the affinity of the complementary arc, and Eq.~\eqref{eq:bound-ep} returns $j_c \mathcal{A}_B+j_c \mathcal{A}_A=j_c \mathcal{A}_c=\sigma$. If instead $A$ is split into disconnected blocks, $B$ acquires several components with distinct arc affinities, so excursions crossing different components then dissipate differently at the same value of $q$, and therefore the bound no longer saturates.

\section{Experimental details}
\label{appendix:experiment}

\begin{figure}[h]
    \centering
    \includegraphics[scale=1.0]{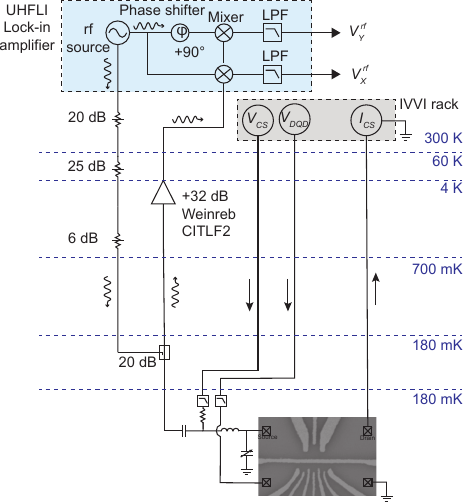}
    \caption{Experimental setup. The device is mounted on a
printed circuit board containing a lumped element radiofrequency (rf) cavity with an integrated bias tee. Gray wavy lines show rf signal
propagation and gray straight lines the direct current (dc) path. The
rf input signal is generated by a Zurich Instruments UHFLI
lock-in amplifier. The reflected signal is amplified at 4K using
a CITLF2 low-noise cryogenic amplifier. Dashed purple lines
indicate the different stages of the dilution refrigerator. Dual
phase demodulation of the reflected rf signal occurs within
the lock-in amplifier (blue shaded region). LPF stands for
low pass filter.}
\label{fig:fridge_setup}
\end{figure}

Figure \ref{fig:fridge_setup} shows the experimental setup. The DQD device is shown by the micrograph and is loaded at the mixing chamber stage of a dilution refrigerator. The charge occupation state of the DQD is probed by two methods via the charge sensing dot, direct current (dc) measurement and radiofrequency (rf) reflectometry \cite{vigneau2023probing}, shown by the two signal paths in the figure. For dc measurement a voltage is applied across the charge sensor denoted by $V_{CS}$. For rf reflectometry a signal is produced by a lock-in amplifier where it is also demodulated and measured on return. The two methods measure nominally identically, in this manuscript the rf measurement data is used for analysis.

\begin{figure}[h]
    \centering
    \includegraphics[scale=0.75]{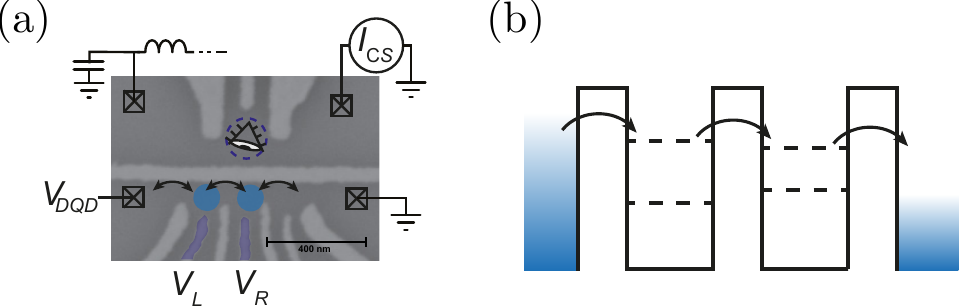}
    \caption{(a) Scanning electron microscope image of a device similar to the one used in this experiment. Blue circles represent the DQD and the eye represents the charge sensor dot used to probe the charge occupation state of the DQD. $V_{L}$ and $V_{R}$ are the voltages applied to gate electrodes that control the number of charges within the left and right dot respectively. The arced arrows represent the possible movement of charges, $I_{CS}$ is the current measured through the charge sensor in dc measurement, $V_{DQD}$ is the source-drain voltage bias across the DQD.
    (b)
    Illustrative energy level representation of the DQD. Energy levels are denoted by black dashed lines. By applying a positive voltage across the DQD transport of charges (curved arrows) is biased from source to drain.}
\label{fig:DQD_setup}
\end{figure}

In particular for rf reflectometry, the reflected signal is demodulated into two quadrature components, $V_X^{\mathrm{rf}}$ and $V_Y^{\mathrm{rf}}$. To improve the separation between charge states, we apply principal component analysis (PCA) to these two channels, independently for each measurement run i.e, each DQD bias value. The first principal component, corresponding to the direction of maximum variance in the two-dimensional sensor signal, is retained as a time trace for subsequent analysis. This procedure combines the information from both quadratures without requiring a prior choice of measurement axis. The three measurement runs used correspond to a DQD bias of 0.855 V and $V_{CS}$ of 0.246 V with an rf frequency of 115 MHz, a DQD bias of -0.955 V and $V_{CS}$ of 0.345 V with an rf frequency of 114 MHz,  a DQD bias of 0.705 V and $V_{CS}$ of 0.246 V and rf frequency of 115 MHz. In all cases the rf power outputted from the lock-in amplifier was -20 dBm. The integration time for the acquisition card, used for both dc and rf, was set to 5 ms.

Figure \ref{fig:DQD_setup} shows the setup at the level of the device, which was operated in the latched regime, where tunnel barrier voltages of the DQD are initially adjusted until all transitions can be sensed and result in a telegraphic signal measured by the charge sensor.
A simplified version of this setup is also shown in Fig.~\ref{fig:markovian-network}(d) of the main text.

\section{Trajectory fitting and transition matrix estimation}
\label{appendix:data-fit}

Two of the results of this Letter are inequalities: quantities computed from excursion data alone are bounds or estimators to quantities defined by the full microscopic dynamics. The verification of our results therefore requires two separate constructions from the same data trace. The left hand sides are the conditional distributions $P(t~| \pm q)$ [Eq.~\eqref{eq:DB-symmetry}], the ratio $P(q)/P(-q)$ [Eq.~\eqref{eq:bound-max-affinity}] and the entropy production estimator $\hat{\sigma}_{\rm exc}$ [Eq.~\eqref{eq:bound-ep}]. Those quantities require only a rule to decide at each instant in time whether the system occupies the resolved region $A$ or not. A model for the unresolved (or hidden) region or transitions rates is not required. On the other hand, the quantities that appear in the right hand sides, $\mathcal{A}_\text{max}$ and $\sigma$, are inaccessible under the inference paradigm by construction, and obtaining them for comparisons requires a full model of the network with all transition rates. We emphasize that this Appendix section deals with this comparison, but if we were handed only the data trace without any microscopic model, our three inference results would be valid and could still be evaluated, but we would not have an independent number to compare them against. Our bounds and results are therefore robust against experimental difficulties that we describe here.

\subsection{From the sensor signal to a trajectory}

\begin{figure}[h]
    \centering
    \includegraphics[width=0.8\linewidth]{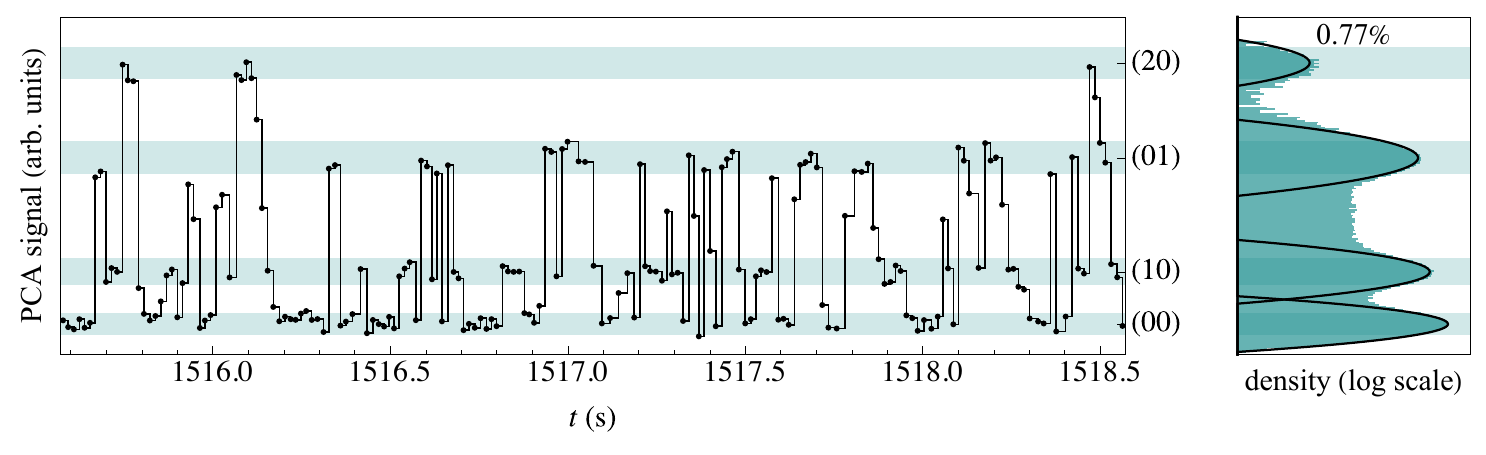}
    \caption{Run 5515 at $0.705$\,V bias. We show a section of the signal with the assigned states and the histogram of the signal over the whole record, logarithmic scale, with the fit of Sec.~\ref{appendix:data-fit}. The fourth component has weight $0.77\%$.}
    \label{fig:fourth-state}
\end{figure}

The analysis uses the PCA signal of Appendix~\ref{appendix:experiment}, i.e. the first principal component of the rf quadratures $V_X^{\rm rf}$ and $V_Y^{\rm rf}$, computed separately for each run and sampled every $\Delta t \simeq 17$ ms ($17.2$ ms at $0.855$ V). Since the principal axis, the sensor operating point and the rf frequency differ between runs, each run is standardized with its own median and median absolute deviation, and the levels below are fitted per run. The overall sign of the principal component is arbitrary; we fix it so that the rare tail lies above level 3.
A sample recorded while the system jumps between two levels averages them over the sampling interval, so it lies between them. We therefore fit to the signal, by expectation maximization, a model with a Gaussian for each level, and for each pair of levels, a component uniform between them and blurred by the noise.
It is started from a three-component Gaussian mixture for the frequently occupied levels $1=(00),\ 2=(10),\ 3=(01)$. State $4$ is too rarely occupied to form a peak of its own (see e.g. Fig.~\ref{fig:markovian-network}(f)), so its level is placed at the densest point among the samples that lie more than $4.5$ standard deviations beyond level $3$, with the width of level $3$; at $0.705$ V, where it forms a peak, it is fitted instead (see Appendix~\ref{appendix:fourth-state}).
For the time trace at the bias $0.855$ V (the one we used in the main text, see Fig.~\ref{fig:db-inference}) the levels sit at $-0.728$, $0.016$, $1.343$ and $2.328$ in standardized units, with widths $0.088$, $0.109$, $0.118$ for levels $1$, $2$, and $3$, and the components for samples recorded during jumps carry approximately $13\%$ of the samples $(2.6 \%$ at $-0.955$ V and $18\%$ at $0.705$ V). The curves in Figs.~\ref{fig:markovian-network}(f),~\ref{fig:fourth-state},~\ref{fig:results-6013}(f) are this fit.

Each sample is assigned to the nearest level of this fit, in units of that level's width. Visits that last a single sample are then discarded, except for state $4$, whose residences last only about two samples on average, and consecutive repeated labels are merged. Every step acts on single samples or on single visits, and none refers to the direction of time, so the procedure commutes with time reversal. The bounds therefore apply to the processed trajectory, and both sides of each comparison below are computed from the same processed trajectory. 

\subsection{Excursion statistics}
The excursion data follow from the trajectory alone, with levels obtained as described above. Since we emulate that the observer has access only to the right dot, and to the barrier through which an electron enters or leaves it, we choose $A=\{3\}$, and an excursion starts with a jump out of $3$ and ends with the next jump into $3$. For the counting observable that counts $+1$ for $1\to3$ and $-1$ for $3\to1$, the run at $0.855$\,V contains $6424$ excursions in $1915$\,s, of which $3680$ have $q=0$, $2254$ have $q=+1$ and $490$ have $q=-1$. The excursion rate is $K=6424/1915\,\mathrm{s}\simeq3.354\,\mathrm{s}^{-1}$, which gives $\hat{\sigma}_{\rm exc}\simeq1.41\,\mathrm{s}^{-1}$. Moreover, since region $A=\{3\}$ has a single state, every return to $A$ is a renewal point~\cite{cox1967renewal}, so what happens after it is independent of what happened before and no memory is carried over. Successive excursions are therefore independent and identically distributed (iid). All statistical procedures we apply in this work follow from this fact~\cite{Datta1993, Bertail2006}.

\emph{Detailed balance.} The first test we perform concerns the results in Eq.~\eqref{eq:DB-symmetry}.
The equality $P(+1)=P(-1)$ is tested with an exact two-sided binomial test on $n_+$ out of $n_++n_-$~\cite{casella2021statistical}. The equality $P(t~|+1)=P(t~|-1)$ is tested with the Kolmogorov--Smirnov distance between the two empirical distributions. The asymptotic null distribution assumes continuous data, whereas the durations are multiples of $\Delta t$ due to the sampling rate. The $p$-value is therefore obtained by permutation~\cite{Bailey1981, Cox2009}, reassigning the outcome labels at random among the excursions with $q=\pm1$ (for $m = 2\times10^{4}$ permutations). Under the null hypothesis the outcome label is independent of the duration, so the labels are exchangeable and the test is exact. We report $p=(b+1)/(m+1)$, with $b$ the number of permutations whose distance is at least the observed one~\cite{Phipson2010}.

\emph{Error bars.} In our analysis in the main text (see Fig.~\ref{fig:db-inference}), we considered error bar intervals on $\ln[P(+1)/P(-1)]$ and $\hat{\sigma}_{\rm exc}$. Those are evaluated via bootstrap resampling~\cite{efron1994introduction} of whole excursions, so that the duration and outcome of each excursion stay together. Since excursions are iid blocks of the trajectory, resampling them reproduces its statistics~\cite{Datta1993, Bertail2006}. The bands in Fig.~\ref{fig:db-inference}(b) are obtained the same way.

\emph{Conservative lower limits.} Bootstrap intervals are not reliable when one of the counts is small, as for $n_-=1$ at $-0.955$\,V, which we discuss later in Appendix~\ref{appendix:negative-bias}.
In this case, another statistical method is required to provide a reliable account of the confidence level. 
Here we consider one-sided lower limits at $95\%$ joint confidence, built from exact intervals for binomial and Poisson counts, whose coverage is never below the nominal level~\cite{Brown2001}. 
For $\ln[P(+1)/P(-1)]$ we take the Clopper--Pearson lower limit~\cite{CLOPPER1934} on the fraction $n_+/(n_++n_-)$, which is binomial once $n_++n_-$ is fixed.
For $\hat{\sigma}_{\rm exc}$ we combine Garwood (Poisson) limits~\cite{GARWOOD1936} on the counts of each outcome and on the number of excursions, using the lower limits in the terms that increase $\hat{\sigma}_{\rm exc}$ and the upper limits in those that decrease it. Each of the $k$ limits is taken at level $1-0.05/k$, so the joint level is at least $95\%$ by the Bonferroni inequality~\cite{Dunn1961}.

\subsection{Generator, affinities and entropy production}
The benchmarks $\mathcal{A}_{\max}(q)$ and $\sigma_{\Delta t}$ require all transition rates, which means we have to fit a transition rate matrix. We obtain it by maximum likelihood from the same processed trajectory. The likelihood of a continuous-time trajectory under a generator with off-diagonal rates $\hat{W}_{yx}$ is~\cite{Albert1962, Billingsley1961}
\begin{equation}
    \mathcal{L}(W) = \prod_x \left(\prod_{y \neq x} \hat{W}_{yx}^{\hat{N}_{yx}}\right)e^{-\hat{\Gamma}_x \hat{\tau}_x},
\end{equation}where $\hat{\Gamma}_x = \sum_{y \neq x} \hat{W}_{yx}$, $\hat{N}_{yx}$ is the number of recorded jumps $x\to y$ and $\hat{\tau}_x$ the total time spent in $x$. Setting $\partial \ln{\mathcal{L}(W)}/\partial \hat{W}_{yx} = 0$ gives
\begin{equation}
    \hat{W}_{yx} = \dfrac{\hat{N}_{yx}}{\hat{\tau}_x}, \qquad (y \neq x),
\label{eq:mle}
\end{equation}with the diagonal fixed by $\sum_y \hat W_{yx} = 0$. For the run at $0.855$\,V, in $\mathrm{s}^{-1}$ and with columns labelling the initial state, we obtain
\begin{equation}\label{eq:W-positive}
  \hat{W}_{0.855\,\mathrm{V}}\simeq\begin{pmatrix}
-11.608 & 6.853 & 3.848 & 4.274\\
5.571 & -10.358 & 8.545 & 10.380\\
5.991 & 3.411 & -12.442 & 10.990\\
0.046 & 0.094 & 0.048 & -25.644
\end{pmatrix}.
\end{equation}
The steady state, Eq.~\eqref{eq:ss-solution} below, is $|\pi\rangle\simeq(0.3270,\,0.4009,\,0.2696,\,0.0026)$, so state $4$ is occupied $0.26\%$ of the time. The escape rates are $\hat{\Gamma}\simeq(11.6,\,10.4,\,12.4,\,25.6)\,\mathrm{s}^{-1}$ against $1/\Delta t\simeq58.1\,\mathrm{s}^{-1}$, i.e. between two and six samples per residence. Note that $\hat{\Gamma}_3\pi_3$ reproduces the measured excursion rate $K$ by construction, since the maximum-likelihood rates~\eqref{eq:mle} make the empirical occupations stationary; it is therefore not an independent check.

The entropy production is then obtained by first solving the steady state distribution
\begin{equation}\label{eq:ss-solution}
    \hat{W}| \pi \rangle = 0,
\end{equation}which we then use to evaluate
\begin{equation}\label{eq:ep-appendix}
    \sigma_{\Delta t} = \sum_{x\neq y}\hat{W}_{xy} \pi_y \ln{\dfrac{\hat{W}_{xy} \pi_y}{\hat{W}_{yx} \pi_x}} \simeq 1.472\,\mathrm{s}^{-1}.
\end{equation}This is the dashed line in Fig.~\ref{fig:db-inference}(d).

The four-state network has seven cycles. Their affinities, oriented so that they are positive, are listed in Table~\ref{tab:cycles} together with the outcome of the counting observable along each oriented cycle. The largest affinity among the cycles with $q=+1$ is $\mathcal{A}_{\max}(+1)\simeq1.568$, carried by $1\to3\to2\to1$. This is the dashed line in Fig.~\ref{fig:db-inference}(c). The cycles with larger affinity, $2\to4\to3\to2$ and $1\to4\to3\to2\to1$, do not cross the edge $1$--$3$ and therefore are invisible to this observable and do not contribute to our bound in Eq.~\eqref{eq:bound-max-affinity}.

\begin{table}[h]
\centering
\begin{tabular}{lcc@{\hspace{2em}}lcc}
\hline\hline
\multicolumn{3}{c}{$0.855\,$V, ~$\nu$ on $1\to3$} & \multicolumn{3}{c}{$-0.955\,$V, ~$\nu$ on $4\to3$} \\
cycle & $\mathcal{A}_c$ & $q$ & cycle & $\mathcal{A}_c$ & $q$ \\
\hline
$1\to3\to2\to1$ & 1.568 & $+1$ & $1\to2\to3\to1$ & 0.060 & $0$ \\
$1\to4\to2\to1$ & 0.389 & $0$ & $1\to2\to4\to1$ & 1.106 & $0$ \\
$1\to4\to3\to1$ & 0.457 & $-1$ & $1\to4\to3\to1$ & 3.821 & $+1$ \\
$2\to4\to3\to2$ & 1.637 & $0$ & $2\to4\to3\to2$ & 4.867 & $+1$ \\
$1\to2\to4\to3\to1$ & 0.068 & $-1$ & $1\to2\to4\to3\to1$ & 4.927 & $+1$ \\
$1\to3\to2\to4\to1$ & 1.179 & $+1$ & $1\to3\to2\to4\to1$ & 1.046 & $0$ \\
$1\to4\to3\to2\to1$ & 2.026 & $0$ & $1\to4\to3\to2\to1$ & 3.761 & $+1$ \\
\hline\hline
\end{tabular}
\caption{Affinities of the seven cycles of the four-state network, obtained from the generators fitted at $0.855\,$V [Eq.~\eqref{eq:W-positive}] and $-0.955\,$V [Eq.~\eqref{eq:W-negative}]. At each bias, every cycle is written in the orientation in which its affinity is positive, and $q$ is the outcome of the counting observable along that orientation.}
\label{tab:cycles}
\end{table}

Finally, we note what $\hat{\sigma}_{\mathrm{exc}}$ is being compared against. The generator $\hat{W}$ is itself inferred from the $\Delta t$-sampled trace, and $\sigma_{\Delta t}$ denotes the entropy production rate of $\hat{W}$, Eq.~\eqref{eq:ep-appendix}. It is not the entropy production of the sampled process, and no general inequality relates it to $\sigma$. The guaranteed statement is $\hat{\sigma}_{\rm exc} \le \sigma$, which holds for the sampled trajectory because sampling commutes with time reversal and every sampled visit to $A=\{3\}$ is a renewal point. That $\sigma_{\Delta t}$ also lies below $\sigma$ is observed in every emulated record of the next section, where up to about $40\%$ of the visits are too short to be resolved. The ratio $\hat{\sigma}_{\rm exc}/\sigma_{\Delta t}\simeq0.96$ quoted in the main text therefore compares two quantities degraded by the same finite sampling, and this is the appropriate comparison.

\subsection{Benchmarking our bounds and fit}
\label{appendix:benchmark}

The comparisons of Sec.~\ref{appendix:data-fit} test our results against $\hat{W}$, a generator fitted to the same sampled and processed trace. This is the best an experiment can offer, but $\mathcal{A}_{\max}(q)$ and $\sigma_{\Delta t}$ obtained in this way are estimates themselves, degraded by the finite sampling rate. To test the whole procedure against exact values, we emulate the measurement on networks whose rates are known, and pass each emulated record through the same pipeline as the data. The code is part of the analysis notebook~\cite{data}, and all random numbers follow from a single seed.

As an effort to emulate results similar to the experimental data, we take four states connected all-to-all, as in Fig.~\ref{fig:markovian-network}(c). With the spanning tree $\{1\text{--}2,\,2\text{--}3,\,2\text{--}4\}$, the chords $1\to3$, $3\to4$ and $1\to4$ close the fundamental cycles $1\to3\to2\to1$, $3\to4\to2\to3$ and $1\to4\to2\to1$, with affinities $F_1$, $F_2$ and $F_3$. The rate of the jump $x\to y$ is given by
\begin{equation}\label{eq:benchmark-rates}
    W_{yx} = k_{xy}\, e^{(E_x - E_y)/2 + f_{xy}/2},
\end{equation}
with symmetric kinetic parameters $k_{xy}=k_{yx}$, state energies $E_x$, and $f_{xy}=-f_{yx}=F_c$ on the chord of the cycle $c$ and zero on the edges of the tree. The energies cancel around every cycle, so each fundamental cycle has affinity $F_c$ exactly, while they keep state $4$ rarely occupied, as in the data. We draw $F_c$ uniformly in $[-5,5]$, $k_{xy}$ log-uniformly in $[0.5,10]\,\mathrm{s}^{-1}$, $E_1$, $E_2$ and $E_3$ uniformly in $[-1,1]$ and $E_4$ in $[4,7]$, and keep a network only if it lies in the range of the measured runs, to make sure we represent models similar to the data measured. We require every rate to be below $40\,\mathrm{s}^{-1}$, escape rates of states $1$, $2$ and $3$ between $3$ and $20\,\mathrm{s}^{-1}$, their occupations between $0.10$ and $0.55$, and the occupation of state $4$ between $0.1\%$ and $2\%$. The exact $\mathcal{A}_{\max}(q)$ and $\sigma$ follow from $W$. This is what we call the ``ground truth''.

Each network is simulated for $1915$\,s, which is precisely the length of the record at $0.855$\,V. The detector samples every $\Delta t=17.2$\,ms. Each sample is the level of the occupied state averaged over the preceding interval, plus white Gaussian noise, with the levels $(-0.72,\,0.01,\,1.33,\,2.33)$ and the noise width $0.11$ close to those of the run at $0.855$\,V in standardized units (Sec.~\ref{appendix:data-fit}). Since the level is averaged over the interval, a sample taken during a jump lies between two levels, as in the data.

Each emulated record is turned into a trajectory exactly as described in Sec.~\ref{appendix:data-fit}: we fit the same model to its signal, with level $4$ held at the position given by the rule above, assign each sample to the nearest level, and then discard single-sample visits while repeated labels are merged. From this trajectory we compute the excursion estimators for $A=\{3\}$ and the counting observable on the edge $1$--$3$, oriented along the exact net current, so that $q=+1$ is the jump $1\to3$ or $3\to1$, whichever is favored so we gather more statistics. We also compute $\mathcal{A}_{\max}(q)$ and $\sigma_{\Delta t}$ from the generator $\hat{W}$ fitted to the same trajectory by maximum likelihood, as for the data. As a third estimate, we fit a hidden Markov model with four hidden states and Gaussian emissions to the raw signal with the Baum--Welch algorithm~\cite{Baum1966, Rabiner1989}. It returns a transition matrix over one sampling interval, and the generator follows from its matrix logarithm divided by $\Delta t$, with negative off-diagonal entries, which signal that the matrix is not embeddable, set to zero.

The Gaussian mixture is started from quantiles of the signal, and it can split a level that dominates the record into two components. For the data we checked by inspection that the levels are correctly placed [Fig.~\ref{fig:markovian-network}(f) and Fig.~\ref{fig:results-6013}(f)]. For the emulations this check is performed automatically. If one of the levels $1$, $2$ or $3$ is placed more than three noise widths away from its true value, then we discard this and resample a new model. When state $4$ is too rare to be found by the assignment, it is simply absent from the trajectory, as it would be for an experimentalist, and $\hat{W}$ has three states.

\begin{figure*}
    \centering
    \includegraphics[width=0.8\linewidth]{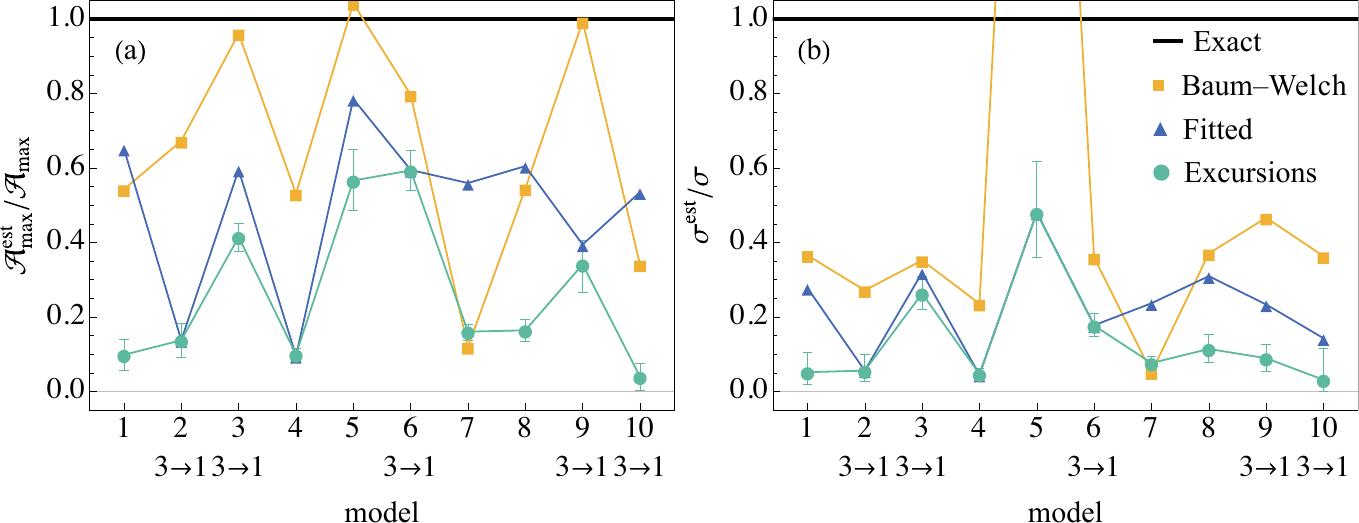}
    \caption{Benchmark of the estimators on ten emulated experiments with a known ground truth. For each network, the black bar is the exact value, the triangles the generator $\hat{W}$ fitted to the processed trajectory (i.e. what we did in the main text), the squares the Baum--Welch fit to the raw signal, and the circles the excursion estimators with $95\%$ bootstrap intervals, for $A=\{3\}$ and the counting observable on the edge $1$--$3$, oriented along the exact net current ($3\to1$ where marked).
    (a) $\ln[P(+1)/P(-1)]$ against $\mathcal{A}_{\max}(+1)$, and the values of $\mathcal{A}_{\max}(+1)$ from the two fitted generators.
    (b) $\hat{\sigma}_{\rm exc}$, $\sigma_{\Delta t}$ and the Baum--Welch value against the exact $\sigma$.
    The Baum--Welch values for network $5$ lies above the plotted range.
    The excursion estimators lie below the exact values in every case.}
    \label{fig:benchmark}
\end{figure*}

Figure~\ref{fig:benchmark} shows the four estimates for ten networks sampled following the approach described above. The excursion estimators lie below the exact values in every case, for $\mathcal{A}_{\max}$ and for $\sigma$. The bounds therefore survive finite sampling, measurement noise and the assignment of the signal to states, none of which enters their derivation.
The estimates $\sigma_{\Delta t}$ fall well below the exact $\sigma$ in every case, and so does the Baum-Welch fit in nine of the ten [see Fig.~\ref{fig:benchmark}], although both use the whole trajectory.
At this sampling rate up to about $40\%$ of the visits are too short to be resolved, and the jumps they contain are lost to every method. This is why the comparison in the main text is made against $\sigma_{\Delta t}$ rather than $\sigma$, both sides are built from the same sampled trace. Additionally, since the real $\sigma$ is not accessible in the experiment, there is no ``ground truth'' to compare against.
Models where $\hat{\sigma}_{\rm exc}$ is much smaller than $\sigma_{\Delta t}$ mean that the dissipation flows mostly through cycles that avoid the edge $1$--$3$, which the observable cannot see.
The Baum--Welch fit is not a lower bound. In network $5$ it exceeds both the exact affinity and the exact $\sigma$. 
Two features of the analysis cause this. First, a sample taken during a jump $1\leftrightarrow 3$ averages the two levels and can land near level $2$. 
The hidden Markov model assumes that each sample reflects the state at the sampling rate, so it reads this sample as a passage $1\to 2\to 3$ that never actually occurred. This adds a spurious directed path around the ring and inflates its affinity. Second, the one-step matrix returned by the fit is not necessarily embeddable. Setting the negative entries of its logarithm to zero can leave transitions with a very small or vanishing reverse rate, and the terms $\ln(W_{xy}/W_{yx})$ that appear in $\sigma$ then become very large. Neither effect can be detected without the ground truth, which an experimentalist does not have. The excursion estimators are inequalities instead, and they stay below the exact values in any network.

\section{Negative bias}
\label{appendix:negative-bias}
To further illustrate our methods and how the inference paradigm can provide new insights into the operation of double quantum dot systems, we consider in this Appendix a different trajectory with negative bias. 
At $-0.955$\,V (run 6013) the same fit and analysis procedure gives $1502$ excursions out of $A=\{3\}$ in total time trace of $1879$\,s. The generator, in $\mathrm{s}^{-1}$ and with columns labelling the initial state, is given by
\begin{equation}\label{eq:W-negative}
  \hat{W}_{-0.955\,\mathrm{V}}\simeq\begin{pmatrix}
-9.750 & 4.452 & 5.067 & 4.400\\
8.033 & -4.911 & 2.128 & 7.333\\
1.704 & 0.421 & -7.200 & 24.199\\
0.013 & 0.037 & 0.005 & -35.932
\end{pmatrix},
\end{equation}
with $|\pi\rangle\simeq(0.3183,\,0.5700,\,0.1110,\,0.0007)$ and escape rates $(9.8,\,4.9,\,7.2,\,35.9)\,\mathrm{s}^{-1}$. The excursion rate is $K\simeq0.7993\,\mathrm{s}^{-1}$, and $\sigma_{\Delta t}\simeq0.083\,\mathrm{s}^{-1}$.

\begin{figure}[h]
    \centering
    \includegraphics[width=0.93\linewidth]{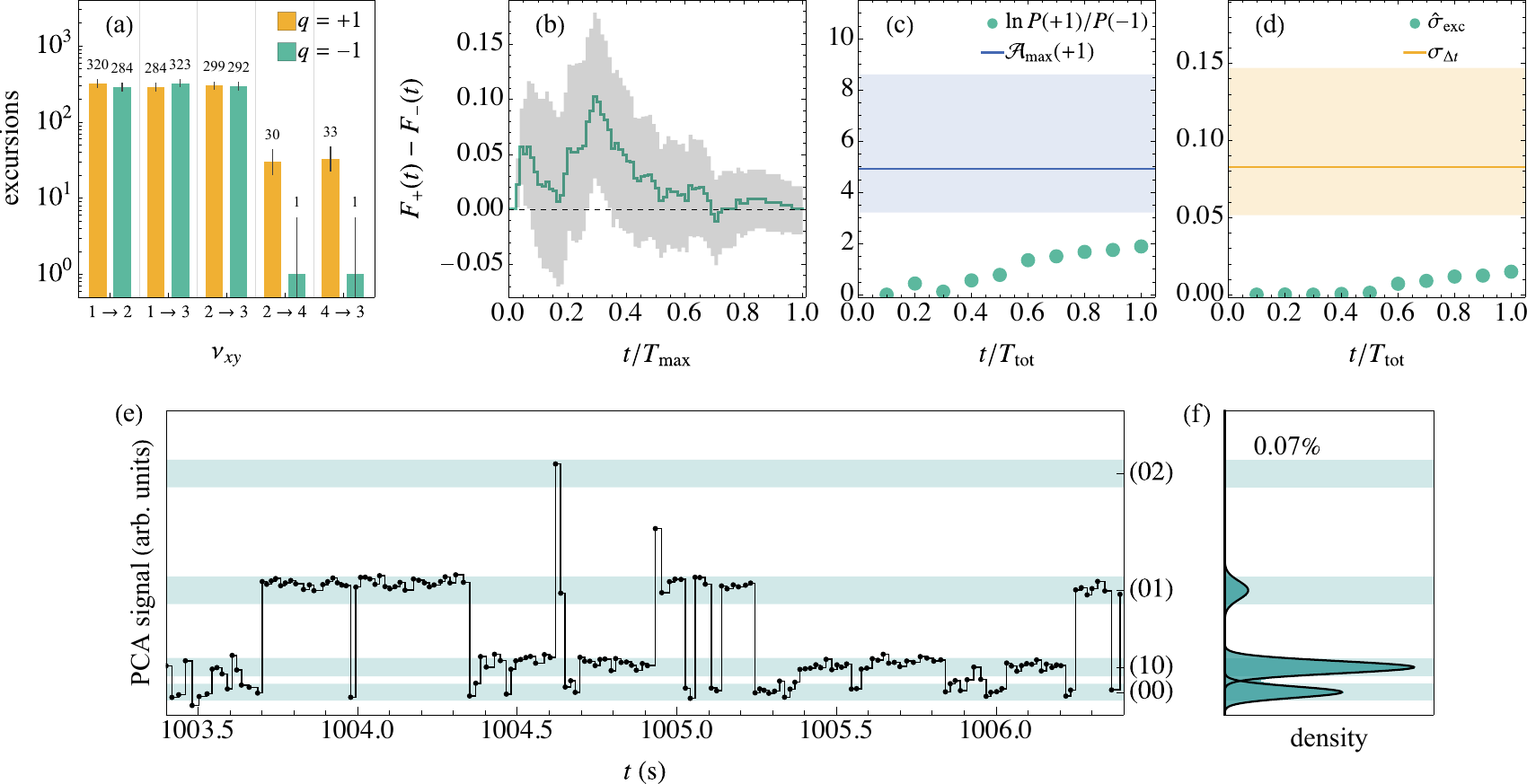}
    \caption{Results of this Letter applied to the double quantum dot at negative bias $-0.955$\,V, with region $A=\{3\}$ (single occupation of the right dot).
    (a) Number of excursions with $q=+1$ and $q=-1$ for counting observables placed on single edges $\nu_{xy}$, with $95\%$ exact Poisson intervals~\cite{GARWOOD1936}. The edges $2\to4$ and $4\to3$, through state $4=(02)$, are crossed almost only in one direction.
    (b) Difference $F_+(t)-F_-(t)$ of the cumulative distributions of the excursion durations for the observable that records $\pm1$ for the jumps $1\leftrightarrow3$, against $t/T_{\max}$ with $T_{\max}$ the $98\%$ quantile of the durations. The line is the data and the shaded region is a $95\%$ bootstrap band. The Kolmogorov--Smirnov distance $D=\max_t|F_+(t)-F_-(t)|\simeq0.11$ has permutation $p$-value $0.048$.
    (c) Conservative $95\%$ lower limit on $\ln[P(+1)/P(-1)]$ for the observable that records $\pm1$ for the jumps $4\leftrightarrow3$, computed from cumulative time fractions $t/T_{\rm tot}$ of the total time record, against $\mathcal{A}_{\max}(+1)\simeq4.93$ (solid line), obtained independently from the generator fitted to the same trace. The shaded band is the range of $\mathcal{A}_{\max}(+1)$ when $\hat{W}_{43}$ is varied over its $95\%$ Poisson interval.
    (d) Conservative $95\%$ lower limit on the excursion estimator $\hat{\sigma}_{\rm exc}$, Eq.~\eqref{eq:bound-ep}, for the same observable and the same cumulative time fractions of the record, against $\sigma_{\Delta t}\simeq0.083\,\mathrm{s}^{-1}$ (solid line) from the fitted generator, with the shaded band obtained in the same way.
    (e) Section of the charge-sensor trace (principal component of the sensor channels) with the assigned states; shaded regions mark the four charge levels, with $4=(02)$ at this bias.
    (f) Histogram of the signal over the whole record with the Gaussian components of the fit of Sec.~\ref{appendix:data-fit}.
    State $4$ is occupied only $0.07\%$ of the time and does not form a separate peak. Both lower limits lie below their benchmarks.}
    \label{fig:results-6013}
\end{figure}

Figure~\ref{fig:results-6013}(a) shows the outcome counts for counting observables placed on single edges: the three edges into state $3$ and, as a control, the edges $1\to2$ and $2\to4$ inside $B$. The edges $1$--$3$ and $2$--$3$ show no significant imbalance, with $284$ against $323$ (binomial $p=0.12$) and $299$ against $292$ ($p=0.81$). For the edge $1$--$3$ the duration test is marginal: the Kolmogorov--Smirnov distance between $P(t\,|+1)$ and $P(t\,|-1)$ has permutation $p$-value $0.048$ [Fig.~\ref{fig:results-6013}(b)], which is not significant once the two tests on this edge are combined. 
An observer monitoring the edge $1$--$3$ at this bias therefore finds no violation.

On the other hand, the edge $4$--$3$ is crossed almost only in one direction: $33$ excursions enter $3$ from $4$ without having left towards $4$, against $1$ for the reverse pattern, with two-sided binomial $p\simeq4\times10^{-9}$. With $n_-=1$ the bootstrap is not reliable, so we quote the conservative lower limits of Sec.~\ref{appendix:data-fit}. At $95\%$ confidence, $\ln[P(+1)/P(-1)]\geq 1.88$ (point estimate $3.50$) and $\hat{\sigma}_{\rm exc}\geq0.015\,\mathrm{s}^{-1}$ (point estimate $0.060\,\mathrm{s}^{-1}$). Both lie below their benchmarks, $\mathcal{A}_{\max}(+1)\simeq4.93$, carried by the cycle $1\to2\to4\to3\to1$, and $\sigma_{\Delta t}\simeq0.083\,\mathrm{s}^{-1}$ [Fig.~\ref{fig:results-6013}(c),(d)]. The binomial $p$-value for this observable drops below $10^{-3}$ after about $960$\,s of data, again a first crossing as in the main text. Note that a single excursion with $q=-1$ does not determine $P(t\,|-1)$, so the duration test carries no further information for this edge.

The benchmarks at this bias are themselves uncertain. Only one jump $3\to4$ is recorded, so the rate $\hat W_{43}$ rests on a single event. Varying it over its $95\%$ Poisson interval moves $\mathcal{A}_{\max}(+1)$ between $3.2$ and $8.6$, and $\sigma_{\Delta t}$ between $0.051$ and $0.147\,\mathrm{s}^{-1}$, which are the shaded bands in Fig.~\ref{fig:results-6013}(c),(d). 
The excursion side needs no such rate, because the two counts $33$ and $1$ fix its lower limits directly.
The imbalances along $1\to2$ ($+36$), $2\to4$ ($+29$) and $1\to3$ ($-39$) are not significant on their own, but they match, within statistics, the net current $32$ along $4\to3$. This is consistent with a single circulation $1\to2\to4\to3\to1$ through state $4$, which is occupied $0.07\%$ of the time and does not form a separate peak in the signal histogram [Fig.~\ref{fig:results-6013}(e),(f)].

\section{The fourth state is physical}

\label{appendix:fourth-state}
In the main text we discussed our results for a trajectory at $0.855$\,V bias and in Appendix~\ref{appendix:negative-bias} a trajectory at $-0.955$\,V. In both cases, state $4$ is occupied less than $0.3\%$ of the time and appears only as a shoulder of level $3$ in the signal histogram. 
This raises the question whether this state truly exists or is just noise within statistical confidence intervals.
To verify that it is a genuine charge state and not an artifact of the assignment, we analyzed a run at a lower bias, $0.705$\,V (run 5515), where it is more often occupied. There a free four-component Gaussian mixture fitted to the signal resolves a fourth level with weight $\sim 0.8\%$ at the position predicted. 
The fit of Sec.~\ref{appendix:data-fit} which also accounts for the samples recorded during jumps, still requires it: removing the Gaussian at level $4$ lowers the log-likelihood by about $600$ ($370$ at bias $-0.955$ V and $80$ at $0.855$ V).

The same analysis applied to this run gives 1755 against 644 excursions for the observable on $1 \to 3$ (Kolmogorov--Smirnov permutation $p \simeq 0.045$), $\ln[P(+1)/P(-1)] \simeq 1.00$ against $\mathcal{A}_{\max}(+1) \simeq 1.218$, and $\hat\sigma_{\rm exc} \simeq 0.59\,{\rm s}^{-1}$ against $\sigma_{\Delta t} \simeq 0.62\,{\rm s}^{-1}$. The entropy production bound is again close to saturation. The affinity bound is less tight, because at this bias $\mathcal{A}_{\max}(+1)$ is carried by the cycle $1 \to 3 \to 2 \to 4 \to 1$ through state 4, which is now occupied more often. Against the affinity $1.027$ of the cycle $1 \to 3 \to 2 \to 1$, which carries most of the excursions with $q=+1$, the ratio reaches $98\%$.

\end{document}

%% file: networks_endmatter.tex
\definecolor{plotgreen}{RGB}{115,194,170}
\colorlet{dkplotgreen}{plotgreen!65!black}

\begin{tikzpicture}[scale=1.0]

\tikzstyle{state}=[circle, draw, fill=plotgreen!20, minimum size=22pt, 
                   inner sep=0pt, font=\large]
\tikzstyle{accessible}=[circle, draw, very thick, fill=plotgreen!45, 
                        minimum size=22pt, inner sep=0pt, font=\large]
\tikzstyle{white}=[circle, draw=none, fill=none, 
                        minimum size=22pt, inner sep=0pt, font=\large]

\node[fill=none, draw=none, text=black, font=\large] (a) at (-4.35, 4.5) {(a)};
\node[fill=none, draw=none, text=black, font=\large] (b) at (-1, 4.5) {(b)};
    
%bath shadow
    \begin{scope}[on background layer]
        \shade[inner color=plotgreen!70, outer color=plotgreen!20, opacity=0.5]
            (4.0,5.0) 
            .. controls (5.5,2.0) and (5.2,1.0) .. (3.0, 0.5) 
            .. controls (-0.2,0.5) and (-1.5,1.0) .. (-1.0, 3.0) 
            .. controls (-1.0,4.0) and (-0.5,4.5) .. (0.0, 5.0) 
            .. controls (1.0,5.0) and (2.0,5.0) .. cycle;
    \end{scope}
    
    \node[accessible] (1) at (0,-0.3) {1};
    \node[accessible]      (2) at (4,-0.3) {2};
    \node[accessible]      (3) at (2,-2) {3};

    \node[state] (4) at (2,4) {};
    \node[state] (5) at (0,2.7) {};
    \node[state]      (6) at (4,2.7) {};
    \node[state] (7) at (2,1.5) {};
    
    \node[white] (8) at (0,4.5) {};
    \node[white] (9) at (4,4.5) {};

    \draw (1) -- (3);
    \draw (1) -- (2);
    \draw (2) -- (3);
    \draw (1) -- (5);
    \draw (1) -- (7);
    \draw (2) -- (6);
    \draw (2) -- (7);
    \draw (4) -- (5);
    \draw (5) -- (6);
    \draw (4) -- (6);
    \draw (5) -- (7);
    \draw (6) -- (7);
    
    \draw[dashed] (5) -- (8);
    \draw[dashed] (6) -- (9);
    \draw[dashed] (4) -- (8);
    \draw[dashed] (4) -- (9);

     \draw[thick, dkplotgreen, very thick, ->] (0.1, 0.3) .. controls (0.1,2.4) and (0.3,2.6) .. (2, 2.6)
                           .. controls (3.7,2.6) and (3.9,2.4) .. (3.9,0.3);
     \draw[thick, dkplotgreen, dashed, very thick, ->] (3.5, -0.2) -- (0.5,-0.2);
     \draw[thick, dkplotgreen, dashed, very thick, ->] (3.5, -0.6) .. controls (2,-1.9) and (2.5,-1.5) .. (2, -1.5)
                           .. controls (1.5, -1.5) and (2,-1.9) .. (0.5, -0.6);

    \draw[->, >=latex', very thick] (2.12, -0.75) arc[start angle=60,end angle=-270,radius=0.3cm];
    \node[state,fill=none,draw=none,text=black, scale = 0.7]  at (2,-1)      (C123)                 {$\mathcal{A}_c$};
  
\begin{scope}[xshift = -3.15 cm, yshift = -1 cm]
    \node[accessible] (1) at (0,0) {1};
    \node[state] (2) at (-1,2) {2};
    \node[state] (4) at (1,2) {4};
    \node[state] (3) at (0,4) {3};
    
    \draw (4) -- (1) -- (2);
    \draw[very thick, color = dkplotgreen, ->, >=latex'] (2) -- (3) -- (4) -- (2);
    
    \draw[->, >=latex', very thick] (0.17, 3) arc[start angle=60,end angle=-270,radius=0.3cm];
    \node[state,fill=none,draw=none,text=black, scale = 0.7]  at (0,2.75)      (C123)                 {$\mathcal{A}$};

    \begin{scope}[on background layer]
        \shade[inner color=plotgreen!70, outer color=plotgreen!20, opacity=0.5]
            (-1.6,2) .. controls (-1.6,3.5) and (-0.5,4.8) .. (0,4.8) 
            .. controls (0.5,4.8) and (1.6,3.5) .. (1.6,2) 
            .. controls (1.6,1) and (-1.6,1) .. cycle;
    \end{scope}
    
\end{scope}

     %\node[font=\large\itshape] at (6.2,-0.5) {B};
    
    %\node (obs) at (2.3, 2.8) {$\{\hat{Q}, \hat{T}\}$};
    %\coordinate (edge89) at ($(5)!0.5!(7)$);
    %\draw[->, gray, thick] (obs) to[bend right=20]  (2);
    %\draw[->, gray, thick] (obs) to[bend left=25] (edge89);
    
\end{tikzpicture}   